\documentclass[a4paper,twocolumn,11pt,amsfonts,amssymb]{quantumarticle}
\pdfoutput=1
\usepackage[utf8]{inputenc}
\usepackage[english]{babel}
\usepackage[T1]{fontenc}
\usepackage{amsmath}
\usepackage{amssymb}
\usepackage{amsfonts}
\usepackage{booktabs}
\usepackage{hyperref}
\usepackage{subcaption}
\usepackage[numbers,sort&compress]{natbib}
\newcommand{\Description}[1]{}

\begin{document}
\title{A Reproducible Software Workflow for Unanchored Approximate MUB Optimization: A Case Study in Dimension Six}
\author{Abdul Fatah}
\email{abdul.fatah@research.atu.ie}
\orcid{0009-0007-5254-8595}
\affiliation{Atlantic Technological University, Ireland}

\author{Ian McLoughlin}
\email{ian.mcloughlin@atu.ie}
\orcid{0000-0003-0424-4849}
\affiliation{Atlantic Technological University, Ireland}

\author{Saim Ghafoor}
\email{saim.ghafoor@atu.ie}
\orcid{0000-0001-8627-5723}
\affiliation{Atlantic Technological University, Ireland}

\maketitle
\begin{abstract}
We present a reproducible, parameter-driven software workflow for optimizing approximate mutually unbiased basis (AMUB) configurations in arbitrary dimensions \(d\) using a Lie-algebra unitary parameterization. The workflow is designed for portable execution across CPU, Apple MPS, CUDA-capable GPU, and HPC backends, using a Taylor-series matrix-exponential layer as an accelerator compatibility pathway. As a dimension-six case study, we optimize unanchored configurations across 100 random seeds for basis counts \(n=3,4,5,6\) in \texttt{complex128} and \texttt{complex64} arithmetic. The workflow recovers exact three-basis configurations, identifies a recurrent four-basis partial-exact hub-and-triangle structure, and finds no near-exact pairs for \(n=5\) or \(n=6\) in the reported campaigns under the primary tolerance.

As a hardware-execution check, we embed the representative \(d=6,n=4\) transition unitaries into three-qubit \(8\times 8\) unitaries and execute the resulting circuits on the 156-qubit Heron processor \texttt{ibm\_marrakesh} using subspace post-selection. The measured QPU pairwise losses are dominated by a hardware and compilation noise floor of approximately \(0.02\)--\(0.08\), associated with compiled circuits averaging 37 native \(\mathrm{CZ}\) gates, which obscures the distinction between classically near-exact and defective pairs. The results provide a reproducible computational framework for exploring AMUB landscapes, together with an initial assessment of the challenges involved in executing optimized dimension-six unitaries on current quantum hardware.
\end{abstract}

\section{Introduction}

Mutually unbiased bases (MUBs) are central objects in quantum information theory, combinatorial design, and finite-dimensional Hilbert space geometry. Two orthonormal bases $B$ and $C$ in $\mathbb{C}^d$ are mutually unbiased~\cite{klappenecker2004ConstructionsMutuallyUnbiased} if
\begin{equation*}
|\langle b, c \rangle|^2 = \frac{1}{d}
\quad \text{for all } b \in B, \; c \in C.
\end{equation*}
A collection of bases is mutually unbiased if every pair of bases in the collection satisfies this condition. Such structures play a fundamental role in quantum state tomography, quantum cryptography~\cite{bennett2014QuantumCryptographyPublic}, and operator theory.

In dimensions that are prime or powers of primes, complete sets of $d+1$ mutually unbiased bases are known to exist~\cite{WOOTTERS1989optimalStateMUB}. In composite dimensions~\cite{McNulty2026mutuallyunbiased}, and in particular in $d=6$, the existence of a complete set of seven MUBs remains a longstanding open problem~\cite{bengtsson2007MUBDandHadamard, horodecki2022FiveOpenProblems}. While numerous analytic and computational approaches have been explored~\cite{brierley2010mutuallyunbiasedbasesdimensions}, no definitive resolution is known. 

Computational searches in dimension six often reduce the search space by fixing one or more bases to canonical representatives, such as the identity, Fourier-type matrices, or prescribed families of complex Hadamard matrices. Fixing a single basis can be interpreted as a gauge choice under the common left-unitary action and is not, by itself, necessarily restrictive. However, searches that impose additional canonical forms or restrict subsequent bases to particular Hadamard families can bias the explored landscape toward selected representatives. This motivates complementary unanchored numerical approaches in which all candidate bases are optimized simultaneously and the resulting pairwise defect geometry is analysed after optimization.

This paper adopts a mathematical-software and computational-physics perspective. We present a generalized, parameter-driven software workflow for optimizing and analysing approximate mutually unbiased basis (AMUB) configurations across dimensions, candidate basis counts, random seeds, numerical precisions, and hardware backends. The aim is not to provide a proof of existence or nonexistence for complete MUB sets in dimension six, but to provide a reproducible computational framework for exploring AMUB landscapes and comparing the structures reached by a specified optimizer, parameterization, precision policy, and seed set.

The workflow uses an unanchored formulation in which all candidate bases are treated symmetrically during optimization. Each candidate basis is parameterized as a unitary matrix through Lie-algebra exponentiation~\cite{elena2000ApproxLieAlgerba},
\[
U_k = \exp(iH_k), \qquad H_k = H_k^\dagger,
\]
where the Hermitian generator \(H_k\) is obtained from an unconstrained trainable complex matrix. This construction enforces the unitary constraint through the forward model rather than through projection steps or penalty terms. The AMUB objective measures the total pairwise deviation from the mutually unbiased condition, while retaining the individual pairwise defects as diagnostic quantities. This pairwise decomposition allows each optimized configuration to be interpreted as a weighted complete graph whose edge weights quantify residual AMUB defect.

The implementation is written in Python using PyTorch and NumPy, with configuration-driven execution, structured artifact export, and reproducible post-processing. Each run records optimized basis matrices, pairwise overlap tensors, pairwise diagnostics, unitarity residuals, generator norms, optimization histories, and run metadata. The main structural and precision-comparison experiments use the CPU pathway with native \texttt{torch.matrix\_exp} as the reference implementation. For accelerator execution, the workflow includes a Taylor-series matrix-exponential compatibility layer, used when native complex matrix exponentiation is unavailable or unsuitable on a given backend. This design separates the reference numerical experiments from backend-specific acceleration and benchmarking.

As a case study, we apply the workflow to the dimension-six problem. We first perform a single-seed validation sweep over \(n=2,\ldots,7\), ranging from the positive-control two-basis case to the formal complete-set target \(n=d+1=7\). We then perform 100-seed campaigns for \(n=3,4,5,6\), which form the main nontrivial transition regime studied in this work. These campaigns allow us to distinguish recurrent structural features from initialization-dependent local basins.

The resulting dimension-six experiments show a structured progression. For \(n=3\), the workflow recovers exact three-basis configurations, while also revealing defective local basins for some seeds. For \(n=4\), the workflow repeatedly identifies a partial-exact hub-and-triangle structure: three basis pairs are near-exact, while the remaining three pairs form a localized defective triangle. Thus, the observed four-basis configurations are not exact four-MUB configurations, but structured partial-exact configurations. For \(n=5\) and \(n=6\), no near-exact pairs are observed in the 100-seed campaigns under the primary tolerance, indicating fully defective configurations within the sampled optimization landscape.

The paper also investigates numerical precision by comparing double-precision \texttt{complex128} reference campaigns with reduced-precision \texttt{complex64} campaigns. Both precisions use the same CPU/native-matrix-exponential structural pathway, allowing precision effects to be separated from accelerator-specific Taylor-exponential effects. The comparison focuses on loss values, near-exact pair counts, tolerance sensitivity, basin selection, and unitarity residuals. The qualitative transition to fully defective observed configurations for \(n\geq 5\) is stable across precisions, while near-exact classifications and basin selection are more sensitive in reduced precision.

Finally, we benchmark the workflow across increasing dimensions and hardware backends. These benchmarks are intended to assess portability and scaling rather than to define the reference structural results. They show that accelerator execution is slower for small dimensions, where fixed launch overheads dominate, but becomes advantageous at larger dimensions as dense linear-algebra workloads increase. In this way, the software supports development on consumer-level hardware, reference CPU campaigns, and scalable execution on HPC resources.

Overall, this work contributes a reproducible computational framework for AMUB optimization, together with a dimension-six case study that exposes exact triples, structured four-basis partial-exact basins, and fully defective observed configurations for \(n=5\) and \(n=6\) under the stated workflow and tolerance policy. The results should be interpreted as reproducible numerical evidence about the optimization landscape sampled by this software, not as a proof of existence or nonexistence of complete MUB configurations in dimension six.

\section{Contributions}

This paper makes the following contributions.

\textbf{Reproducible AMUB software workflow.}
We design and implement a parameter-driven software workflow for optimizing approximate mutually unbiased basis configurations across dimensions, candidate basis counts, random seeds, numerical precisions, and hardware backends. The workflow exports self-contained run artifacts, including optimized basis matrices, pairwise overlap tensors, pairwise diagnostics, unitarity residuals, generator norms, optimization histories, and run metadata. This artifact-based design separates optimization from post-processing and supports independent inspection, regeneration, and extension of the reported experiments.

\textbf{Unanchored Lie-algebra optimization model.}
We formulate the AMUB search as an unanchored optimization problem in which all candidate bases are optimized simultaneously. Each basis is parameterized by Lie-algebra exponentiation,
\[
U_k=\exp(iH_k), \qquad H_k=H_k^\dagger,
\]
with Hermitian generators obtained from unconstrained trainable complex matrices. This enforces unitarity through the forward model rather than through projection steps or penalty terms, while retaining the symmetry of the unanchored formulation during optimization.

\textbf{Pairwise defect diagnostics.}
Beyond recording the aggregate AMUB loss, the workflow retains individual pairwise defects, maximum entrywise deviations from \(1/d\), near-exact classifications under multiple tolerances, and overlap summary statistics. This allows each optimized configuration to be analysed as a weighted complete graph whose edge weights quantify pairwise AMUB defect, making it possible to distinguish exact triples, partial-exact hub structures, localized defective subgraphs, and fully defective observed configurations.

\textbf{Backend-aware matrix-exponential policy.}
The main structural and precision-comparison experiments use the CPU/native-\texttt{matrix\_exp} pathway as the reference implementation. To support accelerator execution when native complex matrix exponentiation is unavailable or unsuitable, the workflow also includes a Taylor-series matrix-exponential compatibility layer. This layer is used for accelerator execution and hardware benchmarking, while the structural AMUB conclusions are based on the CPU reference pathway.

\textbf{Dimension-six validation and multi-seed case study.}
We apply the workflow to dimension six using a single-seed validation sweep over \(n=2,\ldots,7\) and 100-seed campaigns over \(n=3,4,5,6\). The single-seed sweep verifies that the same implementation runs from the positive-control two-basis case to the formal complete-set target \(n=d+1=7\). The 100-seed campaigns provide a reproducible empirical sample of the nonconvex optimization landscape for the main transition regime.

\textbf{Numerical evidence for structured defect in \(d=6\).}
In the complex128 reference campaign, the workflow recovers exact three-basis configurations and repeatedly identifies a structured four-basis partial-exact basin with three near-exact pairs and three defective pairs. For \(n=5\) and \(n=6\), no near-exact pairs are observed in any of the 100-seed complex128 or complex64 campaigns under the primary tolerance. These results provide reproducible numerical evidence about the configurations sampled by the present workflow, not a proof of existence or nonexistence of complete MUB sets in dimension six.

\textbf{Precision-aware analysis.}
We compare complex128 and complex64 campaigns using the same CPU/native-matrix-exponential structural pathway. The comparison shows that the qualitative disappearance of near-exact pairs for \(n=5\) and \(n=6\) is stable across the two precisions, while near-exact classifications and basin selection, especially for \(n=4\), are more sensitive in complex64 arithmetic.

\textbf{Hardware portability and scaling benchmarks.}
We benchmark the workflow across increasing dimensions and hardware backends. On the tested benchmark grid \(d\in\{6,12,24,48,96\}\), GPU execution is slower at small dimensions but becomes advantageous at larger dimensions: the crossover occurs between \(d=24\) and \(d=48\) for complex128 and between \(d=48\) and \(d=96\) for complex64. These benchmarks demonstrate the portability of the same software stack across consumer and HPC-style execution environments.

\textbf{Physical QPU execution check.}
We embed a representative \(d=6,n=4\) optimized AMUB configuration into three-qubit circuits and execute the resulting transition unitaries on \texttt{ibm\_marrakesh}. The measured losses show that compilation overhead and hardware noise obscure the distinction between classically near-exact and defective pairs, providing an initial assessment of the practical challenges of executing dimension-six AMUB unitaries on current QPUs.

\textbf{Reproducible visualization and representative-run selection.}
The workflow produces aggregate summaries and representative-run manifests for visualization. These include representative positive controls, exact or near-exact triples, structured four-basis partial-exact configurations, fully defective observed \(n=5\) and \(n=6\) configurations, and an exploratory \(n=7\) complete-target run. The visualization stage reads saved artifacts rather than rerunning optimization, supporting reproducible figure generation.

\section{Mathematical Formulation}
\label{sec:mathematical-formulation}

This section defines the approximate mutually unbiased basis problem studied in this paper. The main numerical case study is carried out in dimension \(d=6\). We therefore consider collections of \(n\) candidate orthonormal bases in the complex Hilbert space \(\mathbb{C}^6\)~\cite{helmberg2008introHilbertspace}. Each basis is represented by a unitary matrix~\cite{littlewood1977theoryofgroupandmatrix} \(U_k \in U(6)\), whose columns are the basis vectors. The formulation below is written explicitly for \(d=6\), while the software implementation is parameterized by the dimension.

\subsection{Mutual Unbiasedness}

Let $B_i=\{u_{i,1},\ldots,u_{i,6}\}$ and $B_j=\{u_{j,1},\ldots,u_{j,6}\}$ be two orthonormal bases of $\mathbb{C}^6$. The bases are mutually unbiased~\cite{WOOTTERS1989optimalStateMUB} if every vector in one basis has equal squared overlap with every vector in the other basis:
\begin{equation}
    |\langle u_{i,a}, u_{j,b} \rangle|^2 = \frac{1}{6},
    \qquad
    1 \leq a,b \leq 6.
    \label{eq:mub-condition-vector}
\end{equation}
Thus, measurement of a state prepared in one basis gives a uniform probability distribution when expressed in the other basis.

Equivalently, if the bases are represented by unitary matrices $U_i$ and $U_j$, with columns given by the corresponding basis vectors, then the cross-Gram matrix $U_i^\dagger U_j$ satisfies
\begin{equation}
    |(U_i^\dagger U_j)_{ab}|^2 = \frac{1}{6},
    \qquad 1 \le a,b \le 6,
\end{equation}
In other words, all entries of $U_i^\dagger U_j$ have magnitude $1/\sqrt{6}$. Although $U_i^\dagger U_j$ is itself unitary, the mutual unbiasedness condition constrains only its entrywise squared moduli. In matrix form, the condition is
\begin{equation}
    |U_i^\dagger U_j|^2 = \frac{1}{6}\mathbf{1}_{6\times6},
    \label{eq:mub-condition-matrix}
\end{equation}
where the absolute value and square are taken entrywise, and $\mathbf{1}_{6 \times 6}$ denotes the $6 \times 6$ all-ones matrix.

For an exact collection of $n$ mutually unbiased bases, condition~\eqref{eq:mub-condition-matrix} must hold for every pair $1 \leq i < j \leq n$. In this paper, we do not attempt to certify exact existence or nonexistence of complete MUB sets in dimension six. Instead, we use a continuous optimization loss function to study approximate configurations and the pairwise defect structures of low-loss configurations found by the specified numerical workflow.

\subsection{Approximate MUB Loss}

For a pair of candidate bases $U_i$ and $U_j$, define the squared-modulus overlap matrix
\[
M_{ij} = |U_i^\dagger U_j|^2,
\]
where the square is taken entrywise. The pairwise approximate-MUB defect is then defined as the squared Frobenius deviation of \(M_{ij}\) from the uniform target matrix:
\begin{equation}
    \ell_{ij}
    =
    \left\|
    M_{ij}
    -
    \frac{1}{6}\mathbf{1}_{6 \times 6}
    \right\|_F^2
    =
    \left\|
    |U_i^\dagger U_j|^2
    -
    \frac{1}{6}\mathbf{1}_{6 \times 6}
    \right\|_F^2 .
    \label{eq:pairwise-loss}
\end{equation}
Thus, \(\ell_{ij}=0\) if and only if the pair \((U_i,U_j)\) satisfies the mutual unbiasedness condition exactly.

The total $n$-basis loss is the sum of all pairwise defects:
\begin{equation}
    \mathcal{L}_n
    =
    \sum_{1 \leq i < j \leq n}
    \ell_{ij}
    =
    \sum_{1 \leq i < j \leq n}
    \left\|
    |U_i^\dagger U_j|^2
    -
    \frac{1}{6}\mathbf{1}_{6 \times 6}
    \right\|_F^2 .
    \label{eq:total-loss}
\end{equation}
Consequently, $\mathcal{L}_n=0$ corresponds to an exact mutually unbiased collection of $n$ bases. In numerical experiments, nonzero values of $\mathcal{L}_n$ quantify the aggregate residual deviation from mutual unbiasedness.

The loss admits a natural pairwise decomposition, which is important for the analysis in this work. Rather than recording only the scalar value \(\mathcal{L}_n\), the workflow retains the individual quantities \(\ell_{ij}\) for all unordered basis pairs. These pairwise terms reveal whether residual defect is localized in a small number of basis pairs or distributed across the entire configuration. Equivalently, an optimized configuration can be viewed as a weighted complete graph on \(n\) vertices: vertices correspond to bases, and edge weights correspond to pairwise defects \(\ell_{ij}\). This representation is used later to distinguish exact pairs, partial-exact structures, and fully defective observed configurations.

\subsection{Unanchored Optimization}

The optimization problem studied here is unanchored, no basis is fixed in advance to the identity, to a Fourier basis, or to any other canonical representative. Instead, all candidate bases $U_1,\ldots,U_n$ are optimized simultaneously. The unanchored AMUB optimization problem is therefore
\begin{equation}
    \min_{U_1,\ldots,U_n \in U(6)}
    \mathcal{L}_n(U_1,\ldots,U_n).
    \label{eq:unanchored-optimization}
\end{equation}

This formulation treats all candidate bases symmetrically during optimization. The objective is invariant under a common left unitary action,
\[
(U_1,\ldots,U_n) \mapsto (WU_1,\ldots,WU_n), \qquad W \in U(6),
\]
because
\[
(WU_i)^\dagger(WU_j)=U_i^\dagger U_j.
\]
Thus, the unanchored formulation retains a global unitary gauge freedom.

Anchored formulations are often used for classification and for fixed unitary-equivalence freedoms~\cite{durt2010OnMutuallyUnbiasedBases,brierley2009MutuallyUnbiasedBases}, Fixing a single basis, for example \(U_1=I\), can be interpreted as a gauge choice under the common left-unitary symmetry. Such a gauge fixing is not, by itself, necessarily a restriction of the underlying equivalence class. However, practical anchored searches may impose additional structure, such as fixing further bases to prescribed Fourier or Hadamard-family representatives. Those additional choices can restrict the explored numerical landscape.

The unanchored formulation used here avoids imposing such representatives during optimization. All bases are allowed to move simultaneously, and structural features are extracted only after optimization from the recorded pairwise defects. This is particularly useful for the present numerical study, where empirically observed configurations can be analysed through their pairwise defect geometry. In later sections, this viewpoint is used to describe structures such as exact triples, hub-like partial-exact configurations, and fully defective observed configurations. These empirical structures are discussed in the context of known rigidity and nonextendability phenomena for MUB configurations in dimension six~\cite{Grassl2009OnSICPOVMsMUBsd6}.

\subsection{Lie-Algebra Parameterization of Unitary Bases}

To enforce unitarity by construction, each candidate basis is parameterized using Lie-algebra exponentiation~\cite{elena2000ApproxLieAlgerba}. For each basis index $k$, we define
\begin{equation}
    U_k = \exp(iH_k),
    \qquad
    H_k = H_k^\dagger .
    \label{eq:lie-exp-param}
\end{equation}
Since $H_k$ is Hermitian, $iH_k$ is skew-Hermitian, and therefore $\exp(iH_k)$ is unitary in exact arithmetic. This uses the Lie group--Lie algebra relation for the unitary group. The representation is not unique, but it provides a differentiable parameterization that is convenient for unconstrained numerical optimization.

In the implementation, $H_k$ is obtained from an unconstrained trainable complex matrix $A_k \in \mathbb{C}^{d \times d}$ by Hermitian symmetrization:
\begin{equation}
    H_k
    =
    \frac{A_k + A_k^\dagger}{2}.
    \label{eq:hermitian-symmetrization}
\end{equation}
This construction discards the anti-Hermitian component of \(A_k\), ensuring that the generator used in the exponential is Hermitian. The resulting representation is redundant: the complex matrix \(A_k\) contains \(2d^2\) real degrees of freedom, whereas the space of Hermitian \(d\times d\) matrices has real dimension \(d^2\). This redundancy is accepted deliberately because it allows the trainable variables to remain unconstrained complex matrices while the forward model enforces the Hermitian and unitary structure.

The trainable variables are therefore the unconstrained matrices \(A_1,\ldots,A_n\), while the matrices appearing in the AMUB loss are the unitary matrices generated by~\eqref{eq:lie-exp-param}. This separates the unitary constraint from the AMUB objective: the optimizer does not need to enforce orthonormality through projection steps or penalty terms. Instead, deviations from exact unitarity arise only from finite-precision arithmetic and the numerical matrix-exponential pathway, and are monitored separately through the unitarity residuals defined below.

Combining~\eqref{eq:lie-exp-param} and~\eqref{eq:hermitian-symmetrization}, the computational parameterization is
\begin{equation}
    U_k(A_k)
    =
    \exp\left(
    i\frac{A_k + A_k^\dagger}{2}
    \right),
    \qquad
    k=1,\ldots,n.
    \label{eq:computational-param}
\end{equation}
The map $A_k \mapsto U_k(A_k)$ is differentiable through the operations used in the forward model, so gradients of the AMUB loss can be propagated through the matrix exponential using automatic differentiation~\cite{Paszke2017AutomaticDI}.

For the dimension-six case study, the resulting numerical optimization problem can then be written as
\begin{equation}
    \min_{A_1,\ldots,A_n \in \mathbb{C}^{6 \times 6}}
    \mathcal{L}_n
    \left(
    U_1(A_1),\ldots,U_n(A_n)
    \right).
    \label{eq:parametric-optimization}
\end{equation}
Although the optimization variables in~\eqref{eq:parametric-optimization} are unconstrained matrices, the corresponding bases used in the loss are unitary by construction in exact arithmetic.

\subsection{Diagnostics Derived from the Loss}

The total loss $\mathcal{L}_n$ provides a scalar measure of aggregate approximate MUB defect, but it does not by itself describe how the residual error is distributed across the basis pairs. For this reason, the workflow records pairwise diagnostics in addition to the aggregate objective value.

For each unanchored pair $(i,j)$, we compute the squared-modulus overlap matrix
\begin{equation}
    M_{ij}
    =
    |U_i^\dagger U_j|^2 ,
    \label{eq:overlap-matrix}
\end{equation}
where the absolute value and square are taken entrywise. The corresponding pairwise loss is
\begin{equation}
    \ell_{ij}
    =
    \left\|
    M_{ij}
    -
    \frac{1}{6}\mathbf{1}_{6 \times 6}
    \right\|_F^2 ,
    \label{eq:pairwise-loss-diagnostic}
\end{equation}
which is the same pairwise defect used in the total loss~\eqref{eq:total-loss}. We also record the maximum entrywise deviation from the target overlap value,

\begin{equation}
    \delta_{ij}
    =
    \max_{1 \leq a,b \leq 6}
    \left|
    (M_{ij})_{ab}
    -
    \frac{1}{6}
    \right|.
    \label{eq:max-deviation}
\end{equation}
The quantity \(\ell_{ij}\) measures the total squared deviation of a pair, while \(\delta_{ij}\) records the worst entrywise deviation. These two diagnostics are complementary: two pairs may have similar Frobenius defects but different maximum entrywise deviations.

Given a tolerance $\tau>0$, a pair $(i,j)$ is classified as near-exact if
\begin{equation}
    \delta_{ij} < \tau .
    \label{eq:near-exact-classification}
\end{equation}
The primary tolerance used in the reported summaries is \(\tau=10^{-6}\). For precision-sensitivity analysis, especially in reduced precision, near-exact counts are also recomputed using relaxed tolerances such as \(10^{-5}\) and \(10^{-4}\). This separates the recorded numerical quantities \(\ell_{ij}\) and \(\delta_{ij}\) from the tolerance-dependent classification rule.

In addition to pairwise AMUB diagnostics, we monitor numerical unitarity by computing, for each optimized basis,
\begin{equation}
    \epsilon_k
    =
    \|U_k^\dagger U_k - I_6\|_F .
    \label{eq:unitarity-residual}
\end{equation}
Although the Lie-exponential parameterization enforces unitarity in exact arithmetic, \(\epsilon_k\) measures deviations introduced by finite-precision arithmetic and numerical matrix exponentiation. These residuals are used as a consistency check: observed AMUB defects should not be attributed to loss of orthonormality unless the corresponding unitarity residuals are large enough to explain them.

The workflow also records generator-norm diagnostics for the Hermitian matrices \(H_k\). These diagnostics are used to monitor the scale of the Lie-algebra parameters and, when the Taylor-series matrix-exponential compatibility layer is used for accelerator execution, to evaluate the relevant truncation-error bounds under the observed generator norms.

\subsection{Interpretation as Pairwise Defect Geometry}

The pairwise decomposition of \(\mathcal{L}_n\) is central to the interpretation of the numerical results. Each optimized configuration determines a weighted complete graph with vertex set \(\{1,\ldots,n\}\). The vertices represent candidate bases, and the edge weight between vertices \(i\) and \(j\) is the pairwise AMUB defect \(\ell_{ij}\). We refer to this weighted-graph representation as the pairwise defect geometry of the configuration.

In this representation, near-zero edges correspond to exact or near-exact MUB pairs under the chosen tolerance, while nonzero edges indicate defective pairwise relations. This graph viewpoint allows configurations with the same or similar aggregate loss to be distinguished by how their residual defect is distributed. For example, a configuration may contain one basis that forms near-exact pairs with several others, together with a localized defective triangle among the remaining bases. Alternatively, it may exhibit fully distributed defect, with every pair carrying nonzero loss.

This interpretation is particularly useful in the dimension-six experiments reported later. It allows the results to be described not only by scalar loss values, but also by structural patterns in the pairwise defects. In particular, the pairwise defect geometry is used to compare exact three-basis configurations, structured four-basis partial-exact configurations, and fully defective observed configurations for larger candidate basis counts. The same diagnostics also support comparison across numerical precisions, since near-exact classifications may depend on tolerance while the underlying pairwise losses and maximum deviations remain explicitly recorded.

\section{Software Workflow and Precision Policy}
\label{sec:software-workflow}

This section describes the computational workflow used to generate, diagnose, summarize, and benchmark approximate mutually unbiased basis configurations. The workflow is designed as a reproducible mathematical-software artifact rather than as a one-off numerical experiment. Each run records the optimization configuration, numerical precision, backend policy, optimized basis matrices, pairwise overlap matrices, diagnostics, optimization history, and run metadata in machine-readable formats. The experimental protocol built on this workflow is described in Section~\ref{sec:experimental-methodology}, and the resulting numerical evidence is reported in Section~\ref{sec:results}.

The workflow separates four tasks: model construction, numerical optimization, diagnostic artifact generation, and post-processing. This separation is important because tables and figures can be regenerated from saved artifacts without rerunning the optimizer. It also allows the same software stack to be used for single-seed validation sweeps, full multi-seed campaigns, precision-comparison experiments, and hardware-scaling benchmarks.

\subsection{Implementation Overview}

The implementation is written in Python and uses PyTorch for differentiable optimization, NumPy for array storage and post-processing, and command-line execution scripts for reproducibility. PyTorch~\cite{pytorch2019AnImperativeStyle} is used to represent the trainable complex matrices, evaluate the Lie-exponential forward model, compute the AMUB loss, and propagate gradients through the computational graph. NumPy is used for exporting and reloading array-based artifacts such as optimized bases and pairwise overlap tensors. The core model, loss, diagnostic, experiment, and I/O routines are parameterized and reusable as a standard Python module and command-line entry points.

The workflow is organized around a parameterized experiment engine. A run is specified by the dimension \(d\), the number of candidate bases \(n\), the numerical dtype, random seed, optimizer settings, backend policy, matrix-exponential pathway, logging interval, tolerance list, and output directory. These parameters are supplied through configuration files and, where needed, command-line overrides.

For each experiment, the workflow performs the following steps:
\begin{enumerate}
    \item construct an unanchored Lie-exponential AMUB model with \(n\) trainable bases in \(\mathbb{C}^d\);
    \item initialize the random number generators and model parameters according to the run configuration;
    \item optimize the total pairwise AMUB loss using Adam;
    \item track the best loss encountered during optimization and store the corresponding basis configuration;
    \item compute pairwise overlap diagnostics, near-exact classifications, unitarity residuals, and generator-norm diagnostics;
    \item save basis matrices, pairwise overlap matrices, diagnostics, optimization histories, and run metadata in a run-specific artifact directory;
    \item aggregate saved runs across seeds, candidate basis counts, and precisions for later analysis.
\end{enumerate}

Each run directory contains the optimized bases, pairwise overlap tensors, pairwise diagnostics, unitarity residuals, generator norms, run metadata, and optimization history. The artifact format is designed so that post-processing scripts can reload saved runs, recompute summaries, and regenerate figures without rerunning the optimization.

This design allows the same code path to be used for single-run validation, multi-seed campaigns, precision-comparison experiments, and hardware benchmarks. In the dimension-six case study, the implementation supports the validation sweep over \(n=2,\ldots,7\), while the main 100-seed structural campaigns focus on \(n=3,\ldots,6\). The latter range is used for the main transition analysis, whereas \(n=2\) serves as a positive-control case and \(n=7\) is included as an exploratory complete-target run.

\subsection{Backend Policy and Hardware Acceleration via Taylor Series}

The Lie-algebra parameterization requires repeated evaluation of complex matrix exponentials of the form \(\exp(iH_k)\). The workflow therefore treats the matrix-exponential pathway as an explicit backend policy. The main structural AMUB campaigns and precision-comparison experiments use the CPU pathway with PyTorch's native \texttt{torch.matrix\_exp} as the reference implementation. This choice separates the reported structural conclusions from accelerator-specific approximations.

The workflow also supports accelerator execution. In the local Apple Silicon environment used during development, the required complex \texttt{torch.matrix\_exp} operation was not available through the PyTorch MPS execution path. To support accelerator execution in such cases, the software includes a Taylor-series matrix-exponential compatibility layer:
\begin{equation}
    T_N(X) = \sum_{m=0}^{N} \frac{X^m}{m!},
    \qquad
    \exp(X) \approx T_N(X).
    \label{eq:taylor-matrix-exp}
\end{equation}
This layer expresses the matrix exponential using dense matrix multiplication and addition, preserving the active tensor dtype and device. It is used as a backend compatibility mechanism for accelerator execution and hardware benchmarking, not as the reference pathway for the main structural AMUB results.

To quantify the truncation error of the Taylor approximation, we use the standard matrix-exponential remainder estimate in a submultiplicative norm~\cite{Higham2008FunctionsMatrix}. For a skew-Hermitian matrix \(X=iH\), with \(H=H^\dagger\), the order-\(N\) Taylor approximation satisfies
\begin{equation}
    \| \exp(X) - T_N(X) \|_2
    \le
    \frac{\|X\|_2^{N+1}}{(N+1)!} e^{\|X\|_2}.
    \label{eq:truncation-error}
\end{equation}
For \(X=iH\), we have \(\|X\|_2=\|H\|_2\). The trainable generators in the workflow are not explicitly constrained to a principal logarithm branch, so the software records generator spectral norms as diagnostics. These recorded norms allow the truncation bound in~\eqref{eq:truncation-error} to be evaluated using observed generator magnitudes when Taylor-based accelerator execution is used.

The Taylor layer also includes a runtime norm safeguard. It monitors the Frobenius norm of the exponent matrix and issues a warning if this norm exceeds the configured threshold of \(3.0\). The warning indicates that the fixed Taylor order may be insufficient and recommends increasing the Taylor order or adjusting optimization parameters. The Frobenius norm is used as a fast conservative proxy for the spectral norm because
\[
    \|X\|_2 \leq \|X\|_F .
\]
This safeguard does not replace the recorded diagnostics: truncation error, floating-point accumulation, matrix-multiplication roundoff, and backend-specific numerical effects are assessed through generator norms, Taylor-vs-native comparison tests where applicable, and unitarity residuals.

Because \(T_N(iH)\) is not exactly unitary in finite-precision arithmetic, Taylor-based accelerator runs are interpreted as approximate exponential runs. They are accompanied by the same unitarity residual diagnostic
\[
    \|U_k^\dagger U_k - I_d\|_F
\]
used throughout the workflow. Thus, accelerator runs can be benchmarked and inspected without conflating Taylor-approximation effects with the CPU/native reference structural experiments.

In the implemented backend policy, matrix exponentials are routed as follows:
\begin{itemize}
    \item On CPU reference runs, the workflow uses PyTorch's native \texttt{torch.matrix\_exp(1j * H)}.
    \item On accelerator runs where the Taylor pathway is selected or required, the workflow uses the Taylor-series compatibility layer with the configured Taylor order.
\end{itemize}

This policy enables the same unanchored AMUB model, loss function, diagnostics, and artifact format to be used across CPU, Apple Silicon/MPS, CUDA-capable GPU, and HPC execution environments. Hardware-scaling results for the accelerator pathway are reported in Section~\ref{subsec:hardware-benchmarks}.

\subsection{Precision Policy}

The workflow treats numerical precision as an explicit part of the experimental configuration. The main reference structural campaign uses \texttt{complex128} arithmetic. This precision is used as the reference because the analysis depends on small residuals, near-exact pair classifications, pairwise maximum deviations, and unitarity checks.

A corresponding \texttt{complex64} campaign is used as a precision-sensitivity experiment. The purpose of the \texttt{complex64} campaign is not to replace the \texttt{complex128} reference, but to test which qualitative structures persist under reduced complex precision. The precision comparison uses the same structural pathway as the reference campaign: CPU execution with native \texttt{torch.matrix\_exp}. This design isolates floating-point precision effects from accelerator-specific Taylor-exponential effects.

For each precision, the workflow records the same diagnostics: best loss, individual pairwise losses, maximum entrywise deviations, near-exact pair counts, unitarity residuals, generator norms, and optimization histories. The comparison therefore focuses on both scalar outcomes and structural diagnostics, including loss statistics, near-exact pair counts, basin selection, and tolerance sensitivity.

For near-exact pair classification, the primary tolerance used in the reported summaries is
\[
    \tau = 10^{-6}.
\]
Because reduced precision can affect maximum entrywise deviations and near-exact classifications, the workflow also recomputes near-exact counts using relaxed tolerances, including
\[
    10^{-5}
    \quad\text{and}\quad
    10^{-4}.
\]
The recorded quantities \(\ell_{ij}\) and \(\delta_{ij}\) are therefore kept separate from the tolerance-dependent near-exact classification. This separation allows the paper to distinguish changes in the underlying pairwise defects from changes caused by the chosen numerical threshold.

\subsection{Optimization Components}

Each run constructs an unanchored AMUB model consisting of $n$ trainable complex matrices $A_1,\ldots,A_n$. For each basis index $k$, the trainable matrix $A_k$ is converted to a Hermitian generator by
\begin{equation*}
    H_k = \frac{A_k + A_k^\dagger}{2},
\end{equation*}
and the corresponding candidate basis is generated by
\begin{equation*}
    U_k = \exp(iH_k).
\end{equation*}
The AMUB objective is then evaluated over all $\binom{n}{2}$ unordered basis pairs using the pairwise loss defined in Eq.~\eqref{eq:pairwise-loss}. Thus, the number of pairwise AMUB constraints grows quadratically with the number of candidate bases.

The optimization uses the Adam optimizer. The learning rate, initialization scale, step count, logging interval, dtype, backend, random seed, and matrix-exponential pathway are part of the run configuration and are recorded in the run metadata. The hyperparameters used in the reported campaigns are summarized in Table~\ref{tab:hyperparameters} in Section~\ref{sec:experimental-methodology}.

During optimization, the workflow records both the current loss and the best loss encountered so far. When a new best loss is found, the corresponding basis matrices are retained for post-processing and artifact export. The saved diagnostics and summaries are therefore computed from the best recorded configuration rather than from an arbitrary final iterate. This is important because the objective is nonconvex and the optimizer trajectory may fluctuate within a local basin.

The workflow does not claim that the resulting configurations are global minimizers. The AMUB objective is nonconvex, and different random seeds can converge to different local basins. For this reason, the main structural experiments use multi-seed campaigns: the purpose is to identify recurrent configurations, basin diversity, and precision-sensitive behavior under a fixed parameterization, optimizer, seed set, and tolerance policy.

\subsection{Saved Artifacts}

For each run, the workflow writes a self-contained artifact directory. The directory name encodes the dimension, number of bases, numerical dtype, and seed:
\[
\texttt{d\{d\}\_n\{n\}\_\{dtype\}\_seed\{seed\}}.
\]
For example, a complex128 run in dimension six with four candidate bases and seed 1234 is stored in a directory of the form
\[
\texttt{d6\_n4\_complex128\_seed1234}.
\]

Each run directory contains:
\begin{itemize}
    \item \texttt{best\_bases.npy}, storing the optimized basis matrices associated with the best recorded loss;
    \item \texttt{pairwise\_overlaps.npz}, storing the pairwise squared-modulus overlap matrices \( |U_i^\dagger U_j|^2 \) for all unordered pairs \(i<j\);
    \item \texttt{pairwise\_diagnostics.json}, storing pairwise losses, maximum entrywise deviations, near-exact classifications, and overlap statistics;
    \item \texttt{unitarity\_residuals.json}, storing the Frobenius residuals \( \|U_i^\dagger U_i - I\|_F \) used to monitor numerical orthonormality;
    \item \texttt{generator\_norms.json}, storing norm diagnostics for the Hermitian Lie-algebra generators;
    \item \texttt{optimization\_history.csv}, storing the logged optimization trace, including current loss, best loss, step index, and elapsed time;
    \item \texttt{run\_metadata.json}, storing problem parameters, optimizer settings, dtype, backend policy, matrix-exponential pathway, tolerance settings, software versions, runtime information, and diagnostic summaries.
\end{itemize}

The artifact structure separates optimization from analysis. Once a run has been completed, the optimized bases and diagnostics can be reloaded from disk for verification, summarization, and visualization without rerunning the optimizer. This is used by the post-processing workflow to generate campaign-level summaries, including precision-comparison tables, hardware-benchmark summaries, and representative-run manifests. These machine-readable summaries provide the numerical inputs for the tables and figures reported in Section~\ref{sec:results}.

\subsection{Visualization Workflow}

The visualization stage is separated from the optimization stage. Visualization routines load saved overlap matrices, pairwise diagnostics, and campaign summaries from disk rather than recomputing optimized bases. This design ensures that the numerical data used for figures are the same saved artifacts used for tabular summaries.

The workflow produces aggregate visual summaries and representative-configuration visualizations. Aggregate plots summarize loss statistics and near-exact pair counts as functions of the candidate basis count \(n\), dtype, and tolerance policy. These plots are generated from campaign-level summaries rather than from individual optimization traces.

For representative configurations, the workflow uses the saved pairwise diagnostics to construct sorted pairwise-loss spectra. Given a configuration with \(n\) bases, the \(\binom{n}{2}\) pairwise losses \(\ell_{ij}\) are sorted and plotted as a compact spectrum of defect values. This representation makes it possible to distinguish configurations with localized defect from configurations in which every pair is defective.

Full pairwise overlap heatmaps are also generated for selected representative runs. These heatmaps display the matrices \( |U_i^\dagger U_j|^2 \) directly. Because the number of pairwise matrices grows as \(\binom{n}{2}\), full heatmap grids become increasingly large as \(n\) increases. For this reason, the main text emphasizes aggregate summaries and sorted pairwise-loss spectra, while full heatmaps are retained as supporting visualization artifacts.

\subsection{Software Generalization and Multi-Dimensional Scalability}
\label{subsec:software-scalability}

Although the numerical case study in this paper focuses on dimension six, the software architecture is parameterized by the dimension \(d\) and the number of candidate bases \(n\). The same model class can be instantiated with different values of \(d\) and \(n\), subject to the available memory, runtime budget, and backend support. For example, a run in dimension \(10\) with eight candidate bases can be initialized by passing the corresponding parameters to the model:
{\small
\begin{verbatim}
model = UnanchoredAMUBModel(d=10, n_bases=8)
\end{verbatim}
}

The mathematical components of the workflow are also modular. The current implementation uses the AMUB loss defined in Eq.~\eqref{eq:total-loss}, but the same Lie-exponential unitary parameterization and artifact pipeline can be reused for other unitary-constrained search problems after replacing the loss function. Examples of related structures include Quantum Latin Squares~\cite{musto2019QuantumLatinSquares}, Unitary Error Bases~\cite{knill1996NonBinaryUnitary}, and biunitary matrices~\cite{Vicary_2019BiunitaryConstructions}. Such adaptations are outside the scope of the present experiments, but the software organization separates the unitary model from the problem-specific constraint loss.

The benchmark experiments in Section~\ref{subsec:hardware-benchmarks} evaluate scaling on the tested dimension grid
\[
d \in \{6,12,24,48,96\}.
\]
These benchmarks compare CPU/native matrix-exponential execution with accelerator execution using the Taylor-series compatibility pathway. The results show that GPU execution is slower at small dimensions on the tested benchmark grid, where fixed accelerator overheads dominate. At larger tested dimensions, the increased dense linear-algebra workload changes the runtime balance: in the reported benchmark summary, the GPU becomes faster between \(d=24\) and \(d=48\) for complex128, and between \(d=48\) and \(d=96\) for complex64.

These benchmark results are used only to characterize the performance of the implemented workflow on the tested configurations. They do not alter the interpretation of the main structural AMUB experiments, which use the CPU/native matrix-exponential reference pathway. Rather, they demonstrate that the same software design can be executed across local and HPC-style backends, with backend choice treated as an explicit part of the experimental configuration.

\section{Experimental Methodology}
\label{sec:experimental-methodology}

This section describes the numerical experiments used to evaluate the unanchored AMUB workflow in dimension six. The experiments are designed to study how optimized approximate MUB configurations change as the number of candidate bases increases, and to assess how the observed configurations depend on random initialization, numerical precision, and backend policy.

The experimental design has three components. First, a single-seed validation sweep over \(n=2,\ldots,7\) verifies that the same parameterized implementation runs from the positive-control two-basis case to the formal complete-set target \(n=d+1=7\) in dimension six. Second, 100-seed campaigns over \(n=3,4,5,6\) provide a reproducible empirical sample of the nonconvex basins reached by the specified optimizer, initialization scale, step schedule, and tolerance policy. Third, precision-comparison and hardware-benchmark experiments separate structural numerical results from backend-specific acceleration effects.

\subsection{Experimental Goals}

The experiments address three questions.

First, we ask how the optimized AMUB configurations change as the number of candidate bases increases. The single-seed sweep covers \(n=2,\ldots,7\), including both the positive-control two-basis case and the complete-set target \(n=7\) in \(d=6\). The complete-target case is included as an exploratory run only; it is not used by itself to draw statistical conclusions about complete MUB existence or nonexistence in dimension six.

Second, we ask whether the observed structures are stable under random initialization. Because the AMUB objective is nonconvex, one run may converge to a local basin that is not representative of the outcomes reached from other initializations. We therefore perform 100-seed campaigns for \(n=3,4,5,6\), using the same optimizer settings and seed set for the reported complex128 and complex64 campaigns.

Third, we ask whether the qualitative structures observed in the complex128 reference workflow persist under complex64 arithmetic. This precision comparison is designed to distinguish structures that are stable across precision from effects caused by reduced numerical resolution, tolerance-sensitive near-exact classifications, or precision-dependent basin selection.

\subsection{Single-Seed Sweep over Candidate Basis Counts}

As an initial validation and exploratory sweep, the workflow is run for \(n=2,\ldots,7\) using complex128 precision, the CPU backend, PyTorch's native \texttt{torch.matrix\_exp} pathway, and fixed seed \(s=1234\). This sweep serves two purposes. First, it verifies that the same parameterized implementation can handle all candidate-basis counts in the range \(2 \leq n \leq 7\), from a single pair of candidate bases to the formal complete-set target \(n=d+1=7\) in dimension \(d=6\). Second, it produces saved artifacts and representative configurations for preliminary inspection before the multi-seed campaigns.

The case \(n=2\) is included as a positive-control case, since an unbiased pair is known to be attainable and should correspond to a near-zero AMUB loss under successful optimization. The case \(n=7\) is included as an exploratory complete-target run. A single \(n=7\) run is not used to make statistical claims about the existence or nonexistence of complete MUB sets in dimension six.

For each value of \(n\), the workflow constructs \(n\) trainable bases in \(\mathbb{C}^6\), optimizes the AMUB loss, records the best configuration encountered during optimization, computes the diagnostics defined in Section~\ref{sec:mathematical-formulation}, and saves the corresponding artifacts. The number of unordered pairwise AMUB terms grows as
\begin{equation*}
    \binom{n}{2},
\end{equation*}
so the sweep also checks that artifact generation and pairwise diagnostic computation operate correctly across increasing numbers of basis pairs.

The single-seed sweep is not used as a statistical characterization of the nonconvex optimization landscape. Instead, it is used as a validation and exploratory step that motivates the subsequent multi-seed campaigns, where initialization sensitivity, basin diversity, and precision effects are examined more systematically.

\subsection{Multi-Seed Campaigns}

To assess sensitivity to initialization, we perform multi-seed campaigns for
\[
n \in \{3,4,5,6\}.
\]
These values are selected because they cover the main transition regime observed in the workflow: exact or near-exact three-basis configurations, structured four-basis partial-exact configurations, and fully defective observed configurations for larger candidate basis counts. The case \(n=2\) is used as a positive-control case in the single-seed sweep and is not the focus of the multi-seed study. The case \(n=7\) is included in the single-seed sweep as the complete-set target, but the reported multi-seed campaigns focus on \(n=3,\ldots,6\).

For each \(n\) in the multi-seed campaign, the workflow runs 100 independent seeds:
\begin{equation*}
s \in \{0,1,\ldots,99\}.
\end{equation*}
This campaign provides a reproducible empirical sample of the basins reached by the specified unanchored Lie-exponential parameterization, Adam optimizer, initialization scale, step schedule, dtype, and tolerance policy. It should not be interpreted as a complete characterization of the full nonconvex landscape.

For every pair \((n,s)\), the model is reinitialized, optimized, diagnosed, and saved independently. The reported multi-seed summaries include the minimum, median, and maximum best loss over the 100 seeds, together with the minimum, median, and maximum number of near-exact basis pairs observed under the primary tolerance.

\subsection{Optimization Settings}

All structural experiments use the unanchored Lie-exponential parameterization described in Section~\ref{sec:mathematical-formulation}. The trainable variables are unconstrained complex matrices \(A_k\), which are symmetrized to Hermitian matrices \(H_k\) before exponentiation. The AMUB loss is evaluated over all unordered basis pairs.

The optimizer is Adam. The learning rate, initialization scale, step count, logging interval, dtype, backend, matrix-exponential pathway, random seed, and tolerance list are specified by the run configuration and recorded in the run metadata. The optimization history stores the current loss and best loss at regular logging intervals.

\begin{table*}[t]
  \caption{Optimization and campaign-level hyperparameters used in the reported AMUB campaigns. The Taylor order applies only to accelerator-backend timing experiments, not to the CPU/native structural campaigns.}
  \label{tab:hyperparameters}
  \begin{tabular}{ll}
    \toprule
    Hyperparameter & Value / Setting \\
    \midrule
    Optimizer & Adam \\
    Learning rate (\(\eta\)) & \(0.02\) \\
    Initialization scale & \(0.05\) \\
    Step count (for \(n < 5\)) & \(1500\) steps \\
    Step count (for \(n \ge 5\)) & \(2000\) steps \\
    Primary numerical precision & \texttt{complex128} reference campaign \\
    Secondary numerical precision & \texttt{complex64} precision-sensitivity campaign \\
    Structural experiment backend & CPU \\
    Structural matrix-exponential pathway & native \texttt{torch.matrix\_exp} \\
    Campaign seed set (\(s\)) & \(\{0,1,\ldots,99\}\) \\
    Single-seed validation seed (\(s\)) & \(1234\) \\
    Primary near-exact tolerance (\(\tau\)) & \(10^{-6}\) \\
    Additional near-exact tolerances & \(10^{-5}\), \(10^{-4}\) \\
    Taylor expansion order (\(N\)) & \(20\) for accelerator timing experiments \\
    \bottomrule
  \end{tabular}
\end{table*}

During optimization, the workflow tracks the best loss encountered rather than relying only on the final iterate. The basis matrices corresponding to the best observed loss are retained and used for subsequent diagnostics and artifact generation. This is important because the objective is nonconvex and Adam iterates can move within or between local regions during optimization.

As summarized in Table~\ref{tab:hyperparameters}, the reported structural campaigns use \(1500\) steps for \(n<5\) and \(2000\) steps for \(n\geq 5\). These values are part of the fixed experimental protocol used for the reported results.

\subsection{Precision Campaigns}

The complex128 workflow is treated as the reference structural campaign. This precision is used for the main structural analysis because the diagnostics include small pairwise deviations from \(1/6\), near-exact pair classifications, and unitarity residuals.

A corresponding complex64 campaign is run using the same parameterized workflow, the same values of \(n\), the same seed set \(s\in\{0,\ldots,99\}\), and the same optimizer settings. The purpose of the complex64 campaign is not to replace the complex128 reference results, but to evaluate which qualitative structures persist under reduced complex precision.

To isolate numerical precision effects from Taylor-truncation or accelerator-backend effects, the primary structural AMUB experiments and precision-comparison campaigns are executed on the CPU backend using PyTorch's native \texttt{torch.matrix\_exp} implementation. Unless otherwise stated, the precision-comparison results therefore refer to this CPU/native reference pathway. Backend-specific Taylor execution is considered separately in the hardware benchmark experiments in Section~\ref{subsec:hardware-benchmarks}.

The precision comparison focuses on:
\begin{itemize}
    \item best loss statistics over the 100 seeds;
    \item the number of near-exact basis pairs under the primary tolerance;
    \item tolerance sensitivity of near-exact classifications;
    \item basin diversity as reflected in the distribution of best losses;
    \item unitarity residuals;
    \item stability of the observed transition from exact or partial-exact configurations to fully defective observed configurations.
\end{itemize}

\subsection{Near-Exact Pair Classification}

For each optimized configuration, we compute the maximum entrywise deviation
\begin{equation*}
    \delta_{ij}
    =
    \max_{a,b}
    \left|
    \left(|U_i^\dagger U_j|^2\right)_{ab}
    -
    \frac{1}{6}
    \right|
\end{equation*}
for every pair $(i,j)$. Given a tolerance \(\tau>0\), a pair is classified as near-exact when
\begin{equation*}
    \delta_{ij} < \tau .
\end{equation*}

The primary tolerance used in the reported summary tables is
\begin{equation*}
    \tau = 10^{-6}.
\end{equation*}
For precision-sensitivity analysis, especially for \texttt{complex64} runs, the workflow also recomputes near-exact counts using the relaxed tolerances
\begin{equation*}
    10^{-5}
    \quad \text{and} \quad
    10^{-4}.
\end{equation*}
The quantities \(\ell_{ij}\) and \(\delta_{ij}\) are recorded independently of the classification threshold. This separation is important because a change in near-exact pair count may reflect either a change in the underlying overlap defects or a tolerance-sensitive classification near the chosen threshold.

\subsection{Pairwise Structural Diagnostics}

The scalar loss \(\mathcal{L}_n\) is not sufficient to characterize the geometry of an optimized configuration. Two configurations can have similar aggregate losses while distributing their pairwise defects differently across basis pairs. Therefore, the workflow records pairwise diagnostics for every unordered pair \((i,j)\), including:
\begin{itemize}
    \item the pairwise loss contribution \(\ell_{ij}\);
    \item the minimum, maximum, mean, and standard deviation of the entries of \(|U_i^\dagger U_j|^2\);
    \item the maximum absolute deviation \(\delta_{ij}\) from \(1/6\);
    \item near-exact classifications under the configured tolerances.
\end{itemize}

These diagnostics are used to analyse the pairwise defect geometry of each optimized configuration. A defective triangle refers to three bases for which the three pairwise relations among them remain defective. A hub-and-triangle configuration refers to a four-basis pattern in which one basis forms near-exact pairs with the other three, while the remaining three bases form a localized defective triangle. A fully defective observed configuration refers to a saved optimized configuration in which no pair is classified as near-exact under the primary tolerance.

For visualization, the pairwise losses \(\ell_{ij}\) are sorted from largest to smallest, producing a compact spectrum of pairwise defects for each representative configuration. This spectrum records the distribution of edge weights in the complete graph of pairwise defects and provides a concise alternative to displaying every overlap matrix in the main text.

\subsection{Representative Configurations}

After the single-seed sweep and multi-seed campaigns, representative runs are selected for visualization from the saved summaries and diagnostics. The selection is based on recorded metadata and pairwise diagnostics rather than manual inspection of heatmaps. Within each structural category, the lowest-loss available run is selected as the representative. If multiple runs have the same rounded loss within the selection rule, the smallest seed index is used as a deterministic tie-breaker.

The representative set includes, when available:
\begin{itemize}
    \item a positive-control \(n=2\) configuration;
    \item an exact or near-exact \(n=3\) configuration;
    \item a structured \(n=4\) partial-exact hub-and-triangle configuration;
    \item a best observed fully defective \(n=5\) configuration;
    \item a best observed fully defective \(n=6\) configuration;
    \item an exploratory \(n=7\) complete-target configuration from the single-seed sweep;
    \item positive-control configurations in dimensions where complete MUB sets are known to exist, when included in the saved representative manifest.
\end{itemize}

These representatives are used to generate sorted pairwise-loss spectra and supporting heatmap visualizations. The representative-run manifest records the selected run identifier, seed, dtype when available, best loss, number of near-exact pairs, and paths to the saved pairwise diagnostics and overlap matrices. This makes the representative selection reproducible and links each plotted configuration to its saved artifacts.

\subsection{Visualization and Figure Construction}

The visualization stage is separated from the optimization stage. Figure-generation routines load saved overlap matrices, pairwise diagnostics, and campaign-level summaries from disk rather than recomputing optimized bases. This ensures that the numerical data used in the figures are the same saved artifacts used to construct the summary tables.

The main paper uses compact visual summaries rather than full heatmap grids for every configuration. Three types of figures are generated from saved artifacts:
\begin{enumerate}
    \item loss summaries as a function of \(n\) and dtype, with min--median--max ranges over seeds;
    \item near-exact pair count summaries as a function of \(n\), dtype, and tolerance where applicable;
    \item sorted pairwise-loss spectra for representative configurations.
\end{enumerate}

Full pairwise heatmap grids are generated as supporting artifacts. These heatmaps visualize the matrices \(|U_i^\dagger U_j|^2\) directly. The number of pairwise overlap matrices grows as \(\binom{n}{2}\). For example, the complete-target case \(n=7\) contains
\[
\binom{7}{2}=21
\]
pairwise overlap matrices. For this reason, the main text emphasizes compact quantitative summaries and sorted pairwise-loss spectra, while full heatmaps are retained as supporting artifacts.

For each plotted figure, the numerical data used to generate the figure are saved separately from the rendered image when available. This allows figures to be regenerated from saved summaries and diagnostics without rerunning the optimization.

\subsection{Scope and Limitations of the Experiments}

The experiments are numerical and do not constitute a proof of existence or nonexistence of complete MUB sets in dimension six. The optimizer is not guaranteed to find global minima, and the observed configurations depend on the chosen parameterization, optimizer, initialization scale, step schedule, numerical precision, backend pathway, seed set, and near-exact tolerance.

The multi-seed campaigns reported here use 100 independent seeds for each \(n=3,\ldots,6\). These campaigns provide a reproducible empirical sample of the basins reached by the specified workflow. They should not be interpreted as a complete characterization of the full nonconvex optimization landscape.

The \texttt{complex128} results are used as the primary reference for the structural analysis. The \texttt{complex64} results are interpreted as a precision-sensitivity study using the same CPU/native-matrix-exponential structural pathway. Differences between \texttt{complex128} and \texttt{complex64} are therefore treated as evidence of precision-dependent optimization behavior or tolerance-sensitive classification, not as a general ranking of one precision as universally superior.

The reported structural results are also conditioned on the unanchored Lie-exponential parameterization and Adam optimizer used in this workflow. Alternative parameterizations, manifold-optimization methods, learning-rate schedules, initialization distributions, or longer optimization budgets may sample different basins. The hardware benchmark results are likewise specific to the tested benchmark grid, dtype choices, step count, and backend policy; they are used to characterize the implemented workflow rather than to establish hardware-independent performance laws.

\section{Results}
\label{sec:results}

This section reports the numerical results obtained from the unanchored AMUB workflow in dimension six. We first summarize the single-seed validation sweep over $n=2,\ldots,7$, then analyze the statistics of the multi-seed campaigns for $n=3,\ldots,6$ under both double-precision (\texttt{complex128}) and single-precision (\texttt{complex64}) arithmetic. Finally, we examine the pairwise defect geometry via sorted loss spectra and discuss the portability and performance characteristics of the hardware-acceleration layer.

The results presented here are numerical and represent the landscape properties accessible to gradient-based optimization under our unanchored Lie-exponential parameterization. Consistent with standard mathematical caution, they do not constitute a formal proof of the existence or nonexistence of complete MUB sets in dimension six.

subsection{\texorpdfstring{Single-Seed Sweep over \(n=2,\ldots,7\)}{Single-Seed Sweep over n=2,...,7}}
\label{subsec:single-seed-sweep}

As an initial validation and exploratory sweep, the workflow was executed for basis counts \(n=2,\ldots,7\) using the reference \texttt{complex128} configuration on the CPU with native \texttt{torch.matrix\_exp} and fixed validation seed \(s=1234\). The purpose of this sweep was to confirm that the same parameterized workflow runs across candidate basis counts from the positive-control two-basis case to the formal complete-set target \(n=d+1=7\) in dimension six. The results are summarized in Table~\ref{tab:single-seedSweep}.

For \(n=2\), the workflow recovered a mutually unbiased pair to numerical precision, with optimized loss approximately \(1.543\times 10^{-30}\). This case serves as a positive control for the loss function, Lie-exponential parameterization, and diagnostic pipeline. For \(n=3\), the fixed-seed run converged to a defective configuration with total loss approximately \(0.051249218996\) and no near-exact pairs under the primary tolerance \(\tau=10^{-6}\).

For \(n=4\), the fixed-seed run reached the same total loss scale, approximately \(0.051249218996\), but with a different pairwise structure: three of the six basis pairs were classified as near-exact, while the remaining three were defective. This configuration is therefore not an exact four-MUB configuration. Instead, it is a structured partial-exact configuration, consistent with the hub-and-triangle description used later: one basis forms near-exact pairs with three others, while the remaining three pairwise relations carry the residual defect.

For \(n=5\), \(n=6\), and \(n=7\), the fixed-seed runs produced configurations with no near-exact pairs under the primary tolerance. The corresponding losses were approximately \(0.359638494882\), \(0.965932502216\), and \(1.584472133131\), respectively. These single-seed outcomes indicate a progression from a positive-control exact pair, through defective or partial-exact configurations at \(n=3,4\), to fully defective observed configurations for \(n\geq 5\) in this validation sweep. Because this sweep uses only one seed per value of \(n\), it is used only as an exploratory validation step; statistical conclusions are deferred to the multi-seed campaigns.

\begin{table*}[t]
  \caption{Single-seed complex128 sweep over $n=2,\ldots,7$ in dimension six. Near-exact pairs are counted using the tolerance $\tau = 10^{-6}$.}
  \label{tab:single-seedSweep}
  \begin{tabular}{ccccc}
    \toprule
    $n$ & Number of pairs & Best loss & Near-exact pairs & Defective pairs \\
    \midrule
    2 & 1  & $1.543\times 10^{-30}$ & 1 & 0 \\
    3 & 3  & $0.051249218996$ & 0 & 3 \\
    4 & 6  & $0.051249218996$ & 3 & 3 \\
    5 & 10 & $0.359638494882$ & 0 & 10 \\
    6 & 15 & $0.965932502216$ & 0 & 15 \\
    7 & 21 & $1.584472133131$ & 0 & 21 \\
    \bottomrule
  \end{tabular}
\end{table*}
The single-seed sweep confirms that the workflow produces saved artifacts and diagnostics across the full candidate-basis range \(n=2,\ldots,7\). However, a single optimization trajectory can be initialization-dependent. We therefore proceed to 100-seed campaigns for \(n=3,\ldots,6\), where initialization sensitivity and recurrent basin structure can be assessed more systematically.

\subsection{\texorpdfstring{Multi-Seed \texttt{complex128} Campaign}{Multi-Seed complex128 Campaign}}
\label{subsec:complex128-campaign}

To examine initialization sensitivity in the reference precision, we performed a 100-seed \texttt{complex128} campaign for each basis count
\[
n \in \{3,4,5,6\},
\qquad
s \in \{0,1,\ldots,99\}.
\]
All runs in this structural campaign used the CPU backend with native \texttt{torch.matrix\_exp}. The aggregated loss statistics and near-exact pair counts under the primary tolerance \(\tau=10^{-6}\) are summarized in Table~\ref{tab:complex128-multiseed}.

For \(n=3\), the campaign shows both exact and defective outcomes. The minimum and median losses are at numerical zero on the displayed scale, with median loss approximately \(1.01\times 10^{-29}\), and the median near-exact pair count is three. Thus, typical runs in the campaign recover exact MUB triples. However, the maximum loss is \(0.134080\), and the minimum near-exact count is zero, showing that some seeds converge to defective local basins. This confirms that even in the three-basis case, the unanchored objective samples multiple basins under the fixed optimization protocol.

For \(n=4\), the minimum and median best losses are both approximately \(0.051249218996\), and the median near-exact pair count is three. Since a four-basis MUB configuration would require all \(\binom{4}{2}=6\) pairs to be near-exact, these runs do not represent exact four-MUB configurations. Instead, the median behavior is a structured partial-exact configuration with three near-exact pairs and three defective pairs. The maximum loss \(0.294147\) and minimum near-exact count zero indicate that higher-loss fully defective basins are also sampled by some seeds.

For \(n=5\), no near-exact pairs were observed in any of the 100 \texttt{complex128} runs under the primary tolerance. The best loss was \(0.359637\), the median best loss was \(0.439033\), and the maximum best loss was \(0.689537\). Thus, every saved \(n=5\) configuration in this campaign is fully defective under the stated tolerance.

For \(n=6\), the same qualitative pattern is observed. The near-exact count is zero for every seed under the primary tolerance, with minimum and median losses both approximately \(0.868532\) and maximum loss \(1.404517\). These results indicate that, within the specified unanchored Lie-exponential workflow and optimization budget, the sampled \(n=5\) and \(n=6\) configurations contain no near-exact MUB pairs.

\begin{table*}[t]
  \caption{Multi-seed \texttt{complex128} campaign for $n=3,\ldots,6$. The near-exact counts list the min/median/max number of near-exact pairs observed across 100 random seeds, using the primary tolerance $\tau = 10^{-6}$.}
  \label{tab:complex128-multiseed}
  \begin{tabular}{cccc}
    \toprule
    $n$ & Seeds & Loss min/median/max & Near-exact min/med/max \\
    \midrule
    3 & 100 & $0.000000/0.000000/0.134080$ & $0 / 3 / 3$ \\
    4 & 100 & $0.051249/0.051249/0.294147$ & $0 / 3 / 3$ \\
    5 & 100 & $0.359637/0.439033/0.689537$ & $0 / 0 / 0$ \\
    6 & 100 & $0.868532/0.868532/1.404517$ & $0 / 0 / 0$ \\
    \bottomrule
  \end{tabular}
\end{table*}
Overall, the \texttt{complex128} campaign shows a clear empirical progression under the reported workflow: exact triples are recovered for many \(n=3\) seeds, \(n=4\) is dominated by a partial-exact three-edge structure rather than an exact four-MUB configuration, and no near-exact pairs are observed for \(n=5\) or \(n=6\) under the primary tolerance.

\subsection{\texorpdfstring{Precision Sensitivity: \texttt{complex128} versus \texttt{complex64}}{Precision Sensitivity: complex128 versus complex64}}
\label{subsec:precision-sensitivity}

To investigate the role of floating-point precision, we repeated the 100-seed campaigns using \texttt{complex64} arithmetic. The precision-comparison campaigns use the same values of \(n\), the same seed set, the same optimizer settings, and the same CPU/native-\texttt{matrix\_exp} structural pathway as the \texttt{complex128} reference campaign. This isolates floating-point precision effects from accelerator-specific Taylor-exponential effects. Table~\ref{tab:precision-comparison} compares the two precisions.

The comparison shows that the qualitative disappearance of near-exact pairs for \(n=5\) and \(n=6\) is stable across the two precisions. For both \texttt{complex128} and \texttt{complex64}, the near-exact min/median/max counts are \(0/0/0\) for \(n=5\) and \(n=6\). The loss statistics are also similar at the minimum and median levels, although the maximum losses differ between precisions.

For \(n=3\), both precisions recover exact or near-exact triples in the median case. The \texttt{complex128} campaign has median loss \(1.01\times 10^{-29}\), while the \texttt{complex64} campaign has median loss \(7.19\times 10^{-13}\). Both have median near-exact count three, and both have minimum near-exact count zero, indicating that defective seeds occur in both precision settings.

For \(n=4\), the precision dependence is more visible in the near-exact classification. Both precisions reach the same loss scale near \(0.051249\) at the minimum and median. However, the median near-exact count is three in \texttt{complex128} and zero in \texttt{complex64}, while the maximum near-exact count in \texttt{complex64} is still three. This indicates that reduced precision affects either the basin reached, the tolerance-level classification of near-exact edges, or both. The recorded pairwise losses and maximum deviations are therefore used in later analysis to distinguish underlying defect structure from tolerance-sensitive classification.

\begin{table*}[t]
  \caption{Precision comparison for multi-seed AMUB campaigns in dimension six. Near-exact min/med/max use the primary tolerance $\tau = 10^{-6}$ across 100 independent seeds.}
  \label{tab:precision-comparison}
  \begin{tabular}{ccccc}
    \toprule
    $n$ & Dtype & Seeds & Loss min/median/max & Near-exact min/med/max \\
    \midrule
    3 & complex128 & 100 & $0.000000/0.000000/0.134080$ & $0 / 3 / 3$ \\
    3 & complex64  & 100 & $0.000000/0.000000/0.111767$ & $0 / 3 / 3$ \\
    4 & complex128 & 100 & $0.051249/0.051249/0.294147$ & $0 / 3 / 3$ \\
    4 & complex64  & 100 & $0.051249/0.051249/0.409379$ & $0 / 0 / 3$ \\
    5 & complex128 & 100 & $0.359637/0.439033/0.689537$ & $0 / 0 / 0$ \\
    5 & complex64  & 100 & $0.359636/0.439033/0.772279$ & $0 / 0 / 0$ \\
    6 & complex128 & 100 & $0.868532/0.868532/1.404517$ & $0 / 0 / 0$ \\
    6 & complex64  & 100 & $0.868532/0.868532/1.493699$ & $0 / 0 / 0$ \\
    \bottomrule
  \end{tabular}
\end{table*}
Thus, the precision comparison supports two conclusions. First, the absence of near-exact pairs in the observed \(n=5\) and \(n=6\) configurations is stable across \texttt{complex128} and \texttt{complex64} under the primary tolerance. Second, the detailed near-exact classification for \(n=4\) is precision-sensitive, even though the median loss scale remains essentially unchanged.

\subsection{Tolerance Sensitivity of Near-Exact Classification}
\label{subsec:tolerance-sensitivity}

The classification of a pair as near-exact depends on the numerical threshold \(\tau\). For this reason, near-exact pair counts are evaluated not only at the primary tolerance
\[
\tau = 10^{-6},
\]
but also at relaxed tolerances
\[
10^{-5}
\quad\text{and}\quad
10^{-4}.
\]
This tolerance sweep is used to distinguish changes in the recorded pairwise defects from changes caused by the classification threshold.

For the \texttt{complex128} reference campaign, the near-exact classifications are stable across the examined tolerance range in the reported summaries. In particular, exact or near-exact pairs remain classified as such, while the fully defective observed configurations for \(n=5\) and \(n=6\) remain without near-exact pairs under the examined tolerances.

For the \texttt{complex64} campaign, the main tolerance sensitivity occurs at \(n=4\). Under the primary tolerance \(\tau=10^{-6}\), the median near-exact pair count for \(n=4\) is zero, while the maximum is three. This indicates that some \texttt{complex64} runs still reach configurations with three near-exact pairs, but the classification is less stable than in the \texttt{complex128} reference campaign. When relaxed tolerances are applied, some \(n=4\) \texttt{complex64} configurations recover the three near-exact hub edges.

By contrast, the \(n=5\) and \(n=6\) campaigns remain fully defective in the reported tolerance checks: no near-exact pairs are observed under the primary tolerance, and the tolerance sweep is used to verify that this conclusion is not solely an artifact of choosing \(\tau=10^{-6}\). These observations should be interpreted within the fixed workflow, optimizer, seed set, precision policy, and tolerance range used in this study.

\subsection{Pairwise Defect Spectra}
\label{subsec:pairwise-defect-spectra}

\begin{figure}[htbp]
  \centering
  \includegraphics[width=0.95\textwidth]{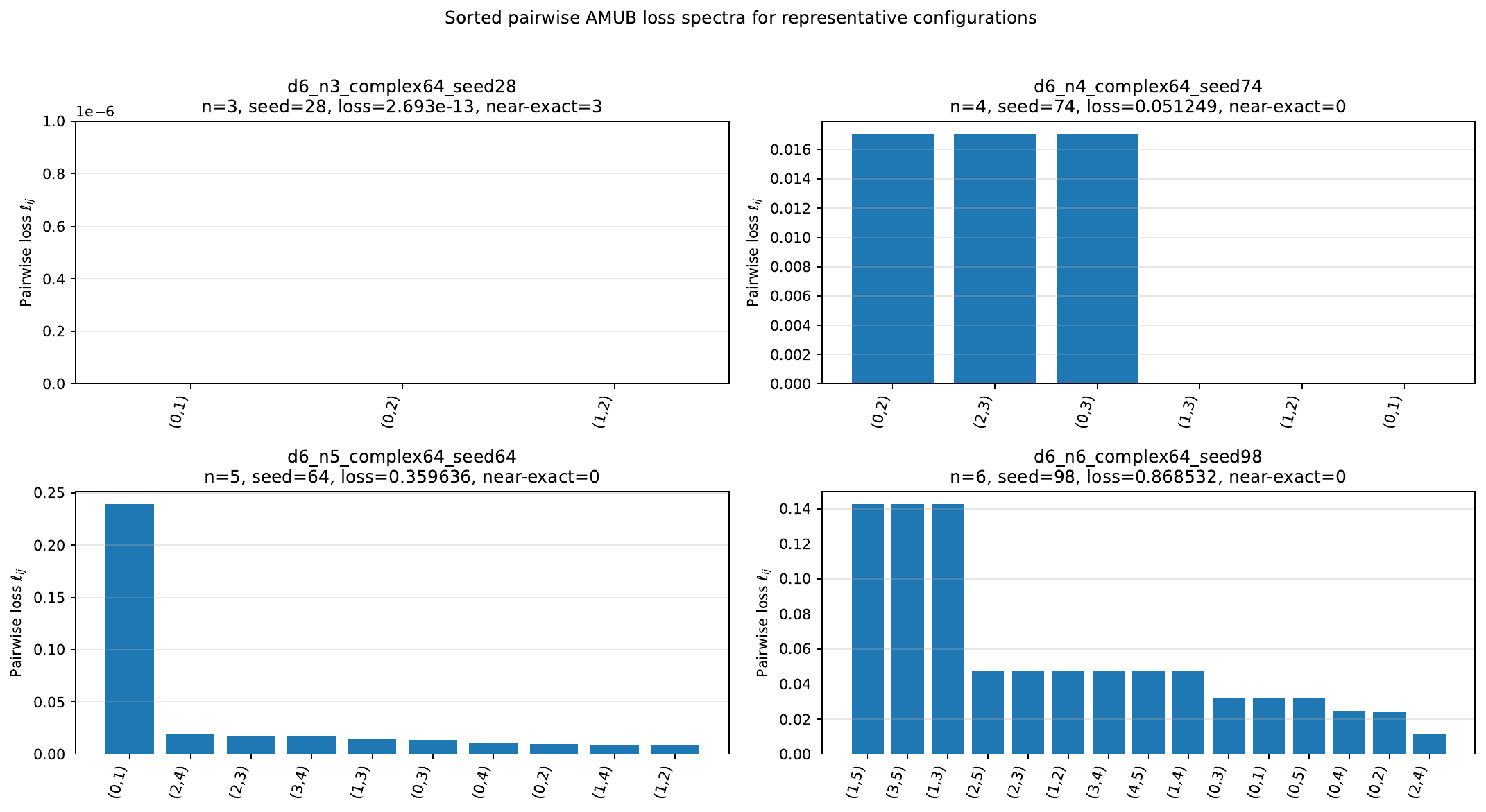}
  \caption{Sorted pairwise AMUB loss spectra for representative complex64 configurations. Machine-zero losses below the display tolerance are shown as zero for visualization only; raw values are retained in the saved figure-data artifact. The figure illustrates that the main transition to fully defective configurations for $n=5$ and $n=6$ persists in complex64, while the sampled basins and near-exact classifications differ from the complex128 reference campaign.}
  \Description{A multi-panel bar chart showing sorted pairwise AMUB loss values for representative complex64 configurations. The n equals 3 exact or near-exact panel has zero displayed bars, the n equals 3 defective panel has three nonzero bars, the n equals 4 panel shows a hub-and-triangle-like structure depending on tolerance, and the n equals 5 and n equals 6 panels have all nonzero bars.}
  \label{fig:pairwise-loss-spectra-complex64}
\end{figure}

The pairwise defect spectra represent the sorted individual pairwise losses \(\ell_{ij}\) for selected representative configurations. They provide a compact structural summary of how the total AMUB defect is distributed across basis pairs. The sorted spectra for representative configurations are shown in Appendix Figures~\ref{fig:pairwise-loss-spectra-complex128} and~\ref{fig:pairwise-loss-spectra-complex64}, where available.

For the exact or near-exact \(n=3\) representative, all three pairwise losses are at numerical zero on the plotted scale. This is consistent with the multi-seed summaries, where the median near-exact count for \(n=3\) is three in both \texttt{complex128} and \texttt{complex64} campaigns.

For the \(n=4\) representative, the spectrum contains three near-zero losses and three nonzero losses. This supports the interpretation of the low-loss \(n=4\) basin as a structured partial-exact configuration rather than an exact four-MUB configuration. In graph terms, one basis acts as a hub forming near-exact pairs with the other three bases, while the remaining three pairwise relations form a defective triangle.

\graphicspath{{figures/}}
\begin{figure}[ht]
  \centering
  \includegraphics[width=1.0\linewidth]{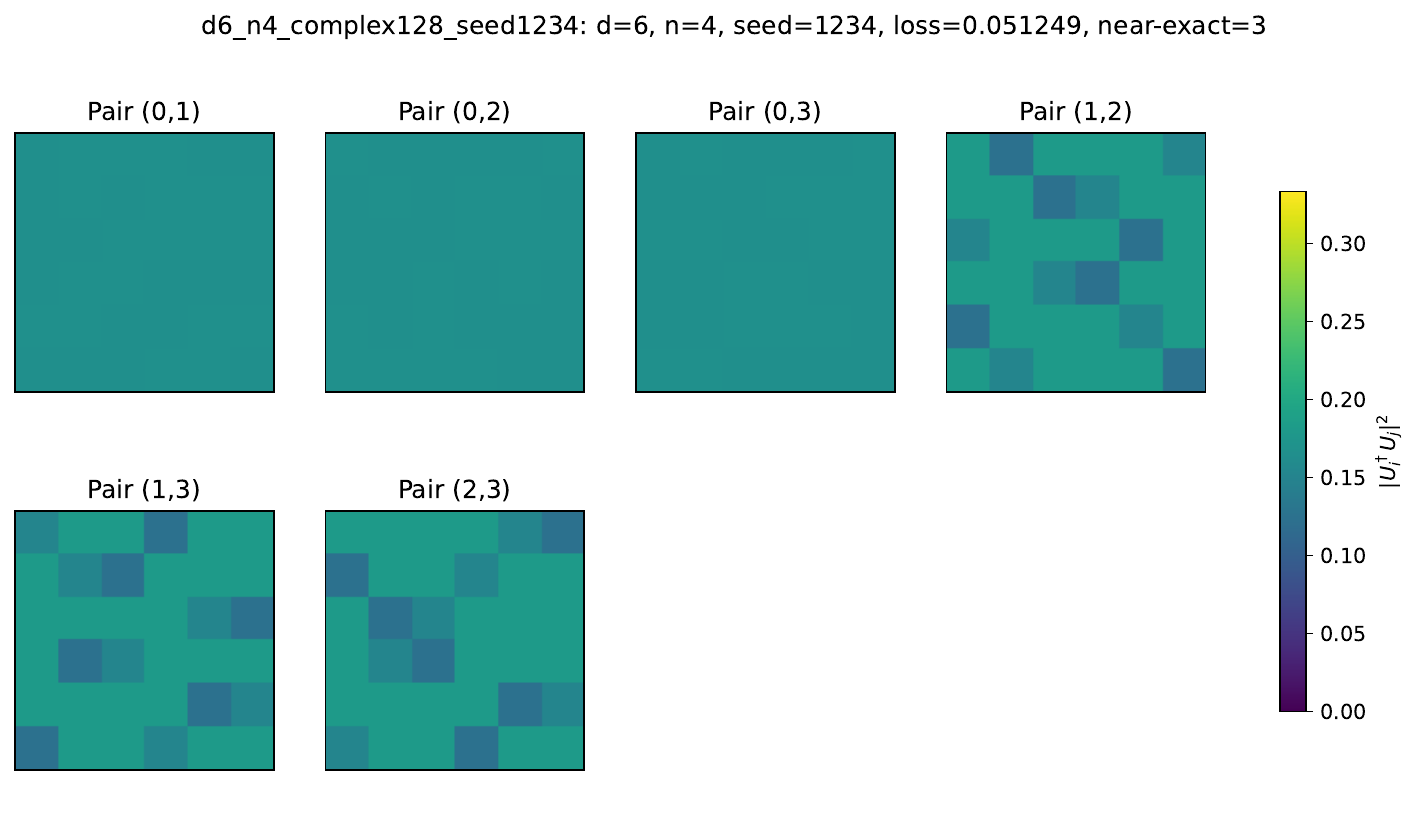}
  \caption{Representative pairwise-loss heatmap for \(d=6\), \(n=4\). The heatmap shows a partial-exact hub-and-triangle structure with three near-zero losses and three nonzero losses.}
  \Description{Heatmap of the pairwise losses for the \(n=4\) representative configuration. Three pairs have near-zero loss, while three pairs have nonzero loss, consistent with a hub-and-triangle pairwise defect structure.}
  \label{fig:heatmap-d6-n4}
\end{figure}

\begin{figure}[ht]
  \centering
  \includegraphics[width=1.0\linewidth]{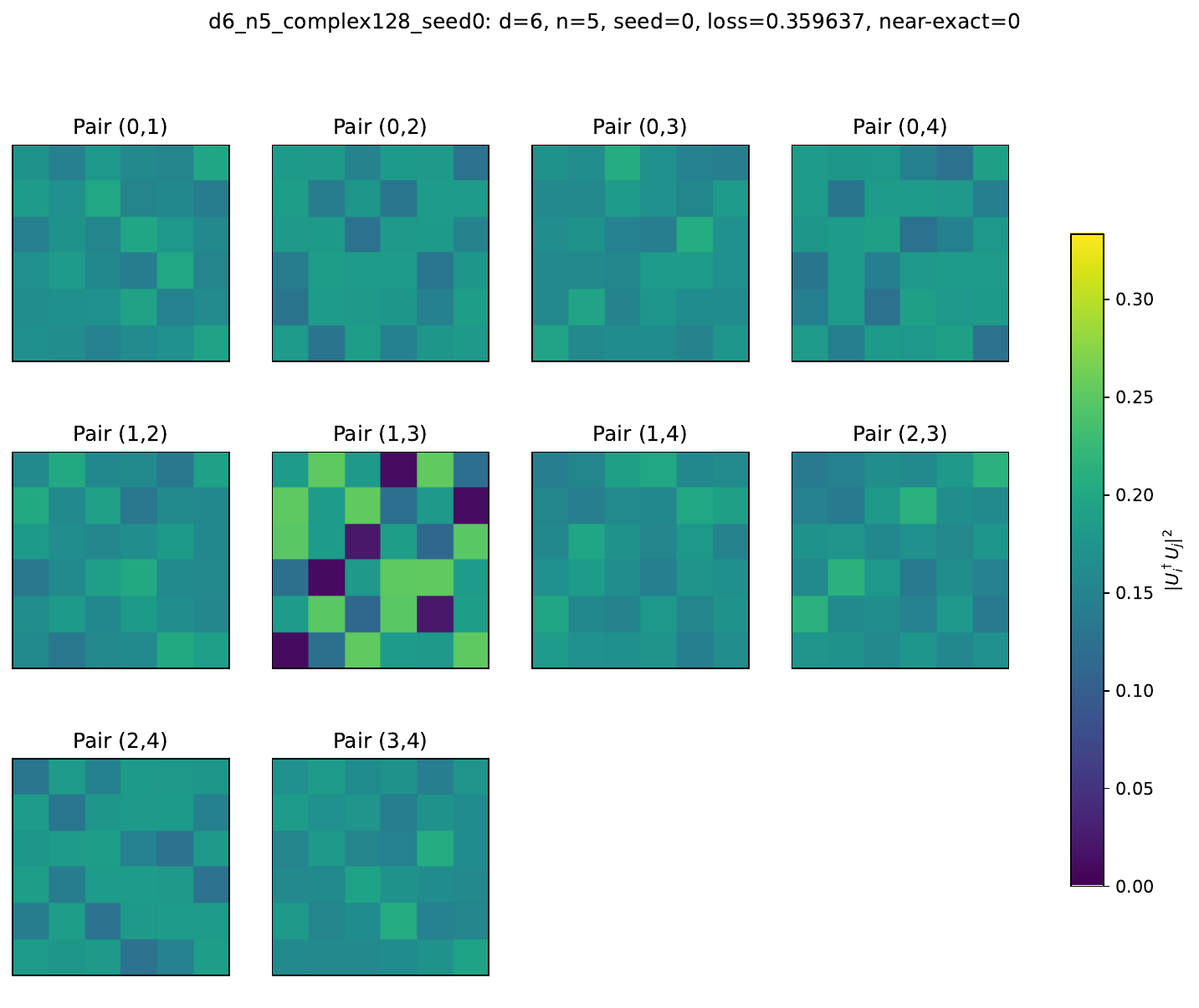}
  \caption{Representative pairwise-loss heatmap for \(d=6\), \(n=5\). The heatmap shows a fully defective observed configuration in which all pairwise losses are nonzero under the primary tolerance.}
  \Description{Heatmap of the pairwise losses for the \(n=5\) representative configuration. All basis pairs have nonzero loss, indicating that no pair is classified as near-exact under the primary tolerance.}
  \label{fig:heatmap-d6-n5}
\end{figure}

For \(n=5\), all ten pairwise losses are nonzero in the selected representative configuration. Thus, the representative \(n=5\) configuration is fully defective under the primary near-exact tolerance. The sorted spectrum shows how the residual defect is distributed across all ten basis pairs rather than localized solely in a small defective subgraph.

For \(n=6\), the selected representative configuration has all fifteen pairwise losses nonzero under the primary tolerance. The corresponding sorted spectrum therefore describes a fully defective observed configuration. The pairwise defect spectrum is useful here because it records not only the absence of near-exact pairs, but also the distribution of defect magnitudes across all basis pairs.

Together, the pairwise spectra support the interpretation suggested by the aggregate summaries: the workflow recovers exact or near-exact triples, finds structured partial-exact four-basis configurations, and samples fully defective observed configurations for \(n=5\) and \(n=6\) under the stated optimization protocol.

\subsection{Physical QPU Execution on IBM Quantum Hardware}
\label{subsec:quantum-verification}

As an additional hardware-execution check, we compiled and executed the representative \(d=6,n=4\) configuration on physical IBM Quantum hardware. The purpose of this experiment is not to use current noisy hardware to certify the fine AMUB defect structure, but to test the physical executability of the optimized transition unitaries and to quantify the effect of compilation and device noise on the measured pairwise AMUB losses.

The representative \(d=6,n=4\) configuration was obtained from seed \(3\). For each basis pair, we formed the classical transition unitary
\[
W_{ij}=U_i^\dagger U_j.
\]
Because \(d=6\) is not a power of two, each \(6\times 6\) transition unitary was embedded into a three-qubit \(8\times 8\) unitary,
\[
V_{ij}=W_{ij}\oplus I_{2\times 2}.
\]
The two additional computational states \(|6\rangle\) and \(|7\rangle\) are therefore outside the active six-dimensional subspace.

For each of the \(\binom{4}{2}=6\) basis pairs, we constructed six circuits corresponding to computational-basis inputs \(|k\rangle\) for \(k\in\{0,\ldots,5\}\). This produced \(36\) circuits in total. The circuits were transpiled using Qiskit's preset pass manager at optimization level \(1\) and executed on the 156-qubit Heron backend \texttt{ibm\_marrakesh} with \(2000\) shots per circuit. IBM's platform lists \texttt{ibm\_marrakesh} as a 156-qubit Heron r2 processor. \cite{IBMQuantumMarrakeshBackend}

To reconstruct the measured transition matrices \(M_{ij}^{\mathrm{QPU}}\), measurement outcomes outside the active subspace, namely bit strings corresponding to \(|6\rangle\) and \(|7\rangle\), were discarded. The remaining counts were normalized over the six-dimensional subspace. This subspace post-selection produces an empirical estimate of the transition probabilities restricted to the embedded \(d=6\) subspace.

Classically, the selected \(d=6,n=4\) representative has a partial-exact hub-and-triangle structure: three pairwise losses are near zero, while the remaining three pairwise losses are nonzero. On the physical backend, the measured pairwise losses were
\[
\ell_{12}^{\mathrm{QPU}}\approx 0.024,\quad
\ell_{23}^{\mathrm{QPU}}\approx 0.044,\quad
\ell_{02}^{\mathrm{QPU}}\approx 0.076
\]
for the classically near-exact pairs, and
\[
\ell_{03}^{\mathrm{QPU}}\approx 0.048,\quad
\ell_{01}^{\mathrm{QPU}}\approx 0.051,\quad
\ell_{13}^{\mathrm{QPU}}\approx 0.070
\]
for the classically defective pairs. Thus, the physical losses lie in an overlapping range for both classes of pair.

The transpiled circuits required nontrivial compilation overhead. In the recorded transpilation data, the circuits averaged 37 native \(\mathrm{CZ}\) gates and approximately 150 single-qubit gates. The resulting physical losses fall in the range \(0.02\)--\(0.08\), which acts as an experimental noise floor for this implementation. This noise floor is larger than the classical distinction between near-exact and defective pair classes in the representative configuration. Consequently, the physical data do not resolve the classical hub-and-triangle structure on the present hardware.

This experiment demonstrates that the optimized AMUB transition unitaries can be embedded, transpiled, executed, and post-selected on a physical QPU, but it also shows that current compilation depth and hardware noise dominate the small classical AMUB defects. The QPU run should therefore be interpreted as a hardware-execution and noise-assessment experiment, not as an experimental proof of the classical AMUB defect geometry.
\begin{figure}[ht]
  \centering
  \includegraphics[width=0.7\linewidth]{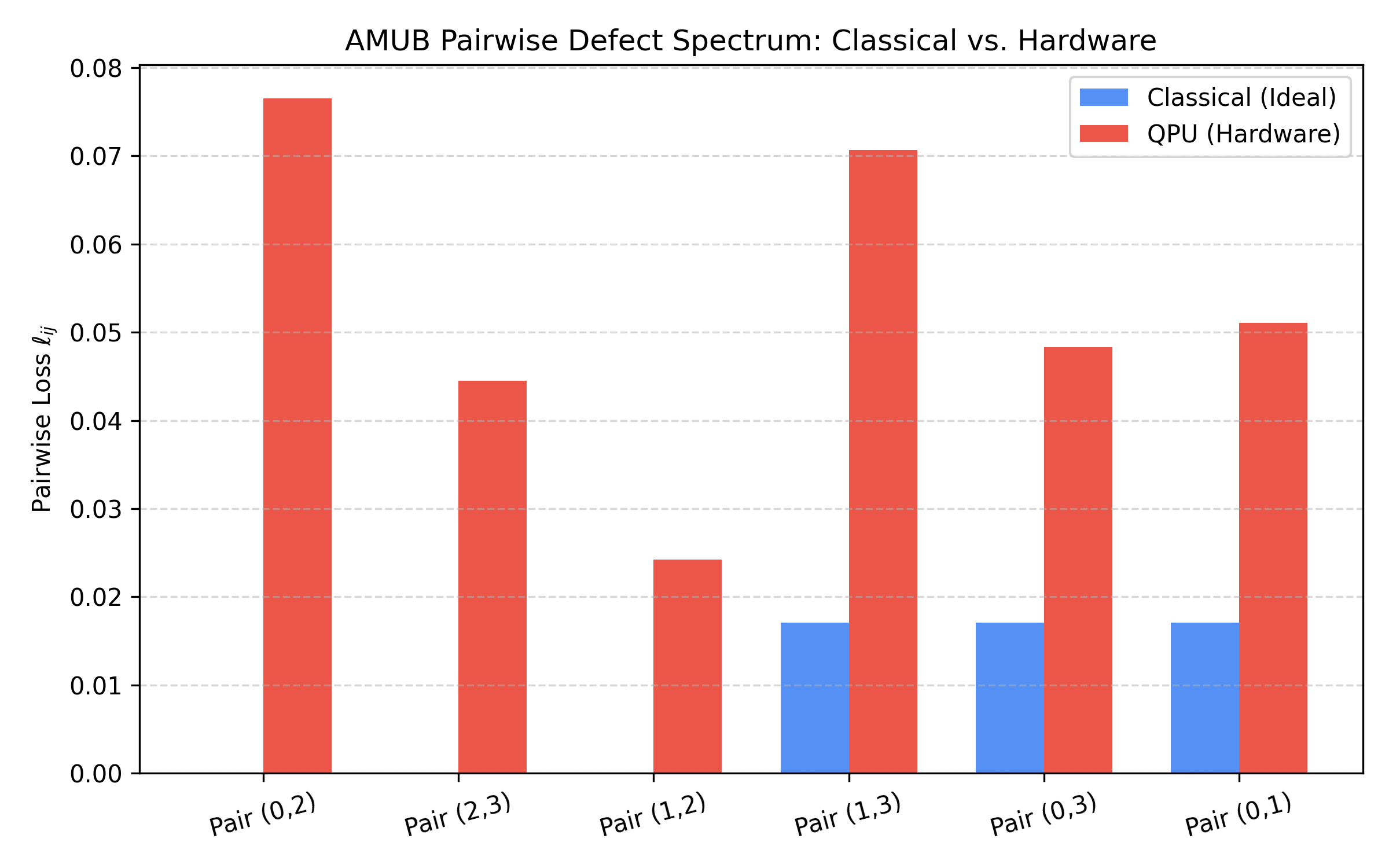}
  \caption{Classical and QPU pairwise-loss comparison for the representative \(d=6,n=4\) configuration. The QPU data were obtained from \(2000\) shots per circuit on \texttt{ibm\_marrakesh} after post-selection onto the active six-dimensional subspace. The measured QPU losses lie in an overlapping range for classically near-exact and defective pairs, indicating that hardware and compilation noise dominate the small classical defect structure.}
  \Description{Bar chart comparing classical and QPU physical pairwise losses for the d=6, n=4 configuration. The QPU losses are all elevated due to physical noise.}
  \label{fig:compare-classical-vs-qpu}
\end{figure}

\subsection{Hardware Benchmarks and GPU Acceleration Performance}
\label{subsec:hardware-benchmarks}

To evaluate backend portability and scaling, we benchmarked the workflow on increasing matrix dimensions. The benchmark uses \(n=6\) candidate bases and measures the wall-clock time required to complete \(1000\) Adam optimization steps for
\[
d \in \{6,12,24,48,96\}.
\]
These timing experiments are separate from the structural AMUB campaigns. The structural campaigns use the CPU/native-\texttt{matrix\_exp} pathway, while the accelerator benchmarks use the Taylor-series compatibility pathway where applicable.

Table~\ref{tab:hardware-benchmarks} reports the workstation benchmark comparing CPU execution against Apple M1 GPU execution through the \texttt{mps} backend. The GPU column uses the order-\(20\) Taylor matrix-exponential pathway.
\begin{table*}[ht]
  \caption{Workstation hardware benchmark in seconds per \(1000\) optimization steps for \(n=6\) bases across dimensions \(d=6,12,24,48,96\). The GPU backend uses the order-\(20\) Taylor matrix-exponential pathway. Relative GPU speed is computed as CPU \texttt{complex128} time divided by MPS GPU time; values above \(1\) indicate GPU speedup relative to the CPU \texttt{complex128} baseline.}
  \label{tab:hardware-benchmarks}
  \begin{tabular}{ccccc}
    \toprule
    Dimension \(d\) & CPU \texttt{c128} (s) & CPU \texttt{c64} (s) & M1 GPU MPS (s) & Relative GPU Speed \\
    \midrule
    6  & 2.617  & 2.695  & 42.958 & \(0.06\times\) \\
    12 & 3.367  & 3.099  & 44.105 & \(0.08\times\) \\
    24 & 4.547  & 4.032  & 43.345 & \(0.10\times\) \\
    48 & 13.077 & 9.782  & 45.735 & \(0.29\times\) \\
    96 & 62.281 & 37.244 & 48.384 & \(1.29\times\) \\
    \bottomrule
  \end{tabular}
\end{table*}

On the workstation benchmark grid, MPS execution is slower than CPU execution for \(d=6,12,24,48\). At \(d=96\), the MPS timing is lower than the CPU \texttt{complex128} timing, giving a relative speed of \(1.29\times\). Therefore, on this tested grid, the workstation crossover occurs between \(d=48\) and \(d=96\). The benchmark does not determine a more precise crossover dimension without additional intermediate dimension points.

To test the same workflow on an HPC-style accelerator backend, we ran the benchmark on the JANUS CUDA backend. The results are summarized in Table~\ref{tab:hardware-benchmarks-janus}. The table reports both \texttt{complex128} and \texttt{complex64} CPU and GPU timings from the saved benchmark summary.

\begin{table*}[ht]
  \caption{JANUS CUDA benchmark in seconds per \(1000\) optimization steps for \(n=6\) bases across dimensions \(d=6,12,24,48,96\). GPU timings use the accelerator matrix-exponential pathway. Speedups are computed as CPU time divided by GPU time for the same dtype; values above \(1\) indicate GPU speedup.}
  \label{tab:hardware-benchmarks-janus}
  \begin{tabular}{cccccc}
    \toprule
    Dimension \(d\) &
    CPU \texttt{c128} (s) &
    GPU \texttt{c128} (s) &
    Speedup \texttt{c128} &
    CPU \texttt{c64} (s) &
    GPU \texttt{c64} (s) \\
    \midrule
    6  & 7.270  & 26.495 & \(0.27\times\) & 7.649  & 26.504 \\
    12 & 10.314 & 27.317 & \(0.38\times\) & 10.870 & 26.522 \\
    24 & 22.282 & 27.453 & \(0.81\times\) & 15.492 & 27.015 \\
    48 & 31.449 & 27.578 & \(1.14\times\) & 22.983 & 27.066 \\
    96 & 72.810 & 27.705 & \(2.63\times\) & 55.595 & 27.054 \\
    \bottomrule
  \end{tabular}
\end{table*}

On the JANUS CUDA benchmark grid, the GPU timing is nearly flat across the tested dimensions, while CPU timing increases with dimension. For \texttt{complex128}, the GPU becomes faster than the CPU between \(d=24\) and \(d=48\). At \(d=96\), the \texttt{complex128} GPU timing is \(27.705\) seconds compared with \(72.810\) seconds on CPU, corresponding to a \(2.63\times\) speedup.

For \texttt{complex64}, the GPU remains slower than CPU at \(d=48\) but is faster at \(d=96\). Thus, on the tested grid, the \texttt{complex64} crossover occurs between \(d=48\) and \(d=96\). At \(d=96\), the \texttt{complex64} GPU timing is \(27.054\) seconds compared with \(55.595\) seconds on CPU, corresponding to approximately a \(2.05\times\) speedup.

These hardware benchmarks show that accelerator execution is not advantageous for small dimensions in this workflow, but becomes beneficial at larger tested dimensions when the dense linear-algebra workload is large enough to offset fixed accelerator overheads. The benchmark results characterize the performance of the implemented workflow on the tested hardware and dimension grid. They are not used to support the structural AMUB conclusions, which are based on the CPU/native reference pathway.

\graphicspath{{figures/}}

\begin{figure}[ht]
  \centering
  \includegraphics[width=1.0\linewidth]{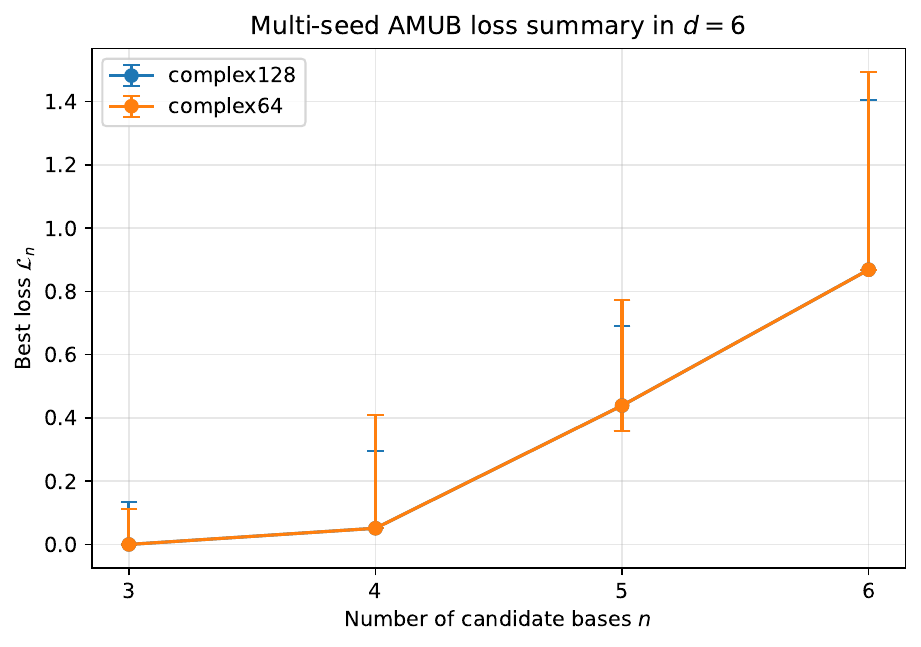}
  \caption{Best AMUB loss summary in \(d=6\). The plot shows median best loss with min--max range across 100 seeds for \texttt{complex128} and \texttt{complex64}.}
  \Description{A line plot with error bars comparing AMUB best loss values for complex128 and complex64 across n equals 3, 4, 5, and 6.}
  \label{fig:loss-summary-by-n-precision1}
\end{figure}

\begin{figure}[t]
  \centering
  \includegraphics[width=1.0\linewidth]{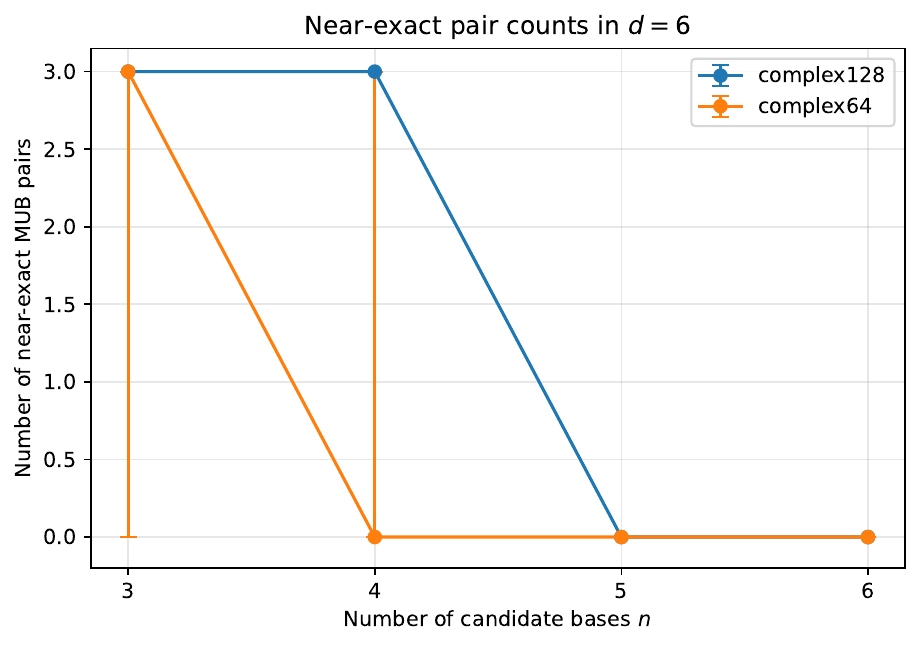}
  \caption{Near-exact MUB pair counts in \(d=6\). The plot shows the median number of near-exact pairs with min--max ranges across 100 seeds. Under the primary tolerance, no near-exact pairs are observed for \(n=5\) or \(n=6\) in either precision.}
  \Description{A line plot with error bars showing the number of near-exact MUB pairs for complex128 and complex64 across n equals 3, 4, 5, and 6.}
  \label{fig:near-exact-pair-summary1}
\end{figure}

\section{Conclusion}

This paper presented a reproducible mathematical-software workflow for unanchored approximate mutually unbiased basis optimization. The workflow is parameterized by dimension, candidate basis count, random seed, numerical precision, and backend policy. Candidate bases are represented through Lie-exponential unitary parameterizations, all bases are optimized simultaneously without imposing a fixed reference basis during optimization, and each run exports machine-readable artifacts, including optimized bases, pairwise overlap matrices, diagnostics, metadata, and optimization histories.

As a dimension-six case study, we applied the workflow to a single-seed validation sweep over \(n=2,\ldots,7\) and to 100-seed \texttt{complex128} and \texttt{complex64} campaigns over \(n=3,\ldots,6\). The results show a reproducible empirical progression under the specified workflow. Exact three-basis configurations are recovered in the reference campaign, while four-basis runs repeatedly identify a structured partial-exact hub-and-triangle configuration with three near-exact pairs and three defective pairs. For \(n=5\) and \(n=6\), no near-exact pairs are observed in any of the reported 100-seed campaigns under the primary tolerance. Thus, within the sampled optimization landscape, the transition from exact or partial-exact pairwise structure to fully defective observed configurations begins at \(n=5\).

The pairwise defect diagnostics show that this transition is not captured by the scalar loss alone. By retaining the individual pairwise losses \(\ell_{ij}\), maximum deviations \(\delta_{ij}\), and near-exact classifications, the workflow allows optimized configurations to be interpreted as weighted complete graphs. This representation distinguishes exact triples, four-basis hub-and-triangle structures, and fully defective observed configurations for \(n=5\) and \(n=6\). The representative-run manifests and sorted pairwise-loss spectra provide a reproducible way to inspect these structures from saved artifacts rather than from manual heatmap selection.

The precision comparison demonstrates that the qualitative disappearance of near-exact pairs for \(n=5\) and \(n=6\) is stable across \texttt{complex128} and \texttt{complex64} in the reported campaigns. At the same time, reduced precision affects detailed basin selection and near-exact classification, most visibly in the \(n=4\) case, where the median near-exact count differs between \texttt{complex128} and \texttt{complex64} under the primary tolerance. This confirms the importance of recording tolerance-dependent classifications separately from the underlying pairwise losses and maximum deviations.

The hardware benchmarks demonstrate that the same software stack can be executed across CPU and accelerator backends. The main structural experiments use the CPU/native-\texttt{matrix\_exp} pathway as the reference implementation, while the Taylor-series matrix-exponential layer provides an accelerator compatibility pathway for timing and portability tests. On the tested JANUS CUDA benchmark grid, GPU execution is slower at small dimensions but becomes faster at larger dimensions, with the crossover occurring between \(d=24\) and \(d=48\) for \texttt{complex128}, and between \(d=48\) and \(d=96\) for \texttt{complex64}. These benchmarks characterize the implemented workflow on the tested hardware and do not alter the interpretation of the CPU-reference structural results.

The results are numerical and do not constitute a proof of existence or nonexistence of complete MUB sets in dimension six. They are conditioned on the unanchored Lie-exponential parameterization, Adam optimizer, initialization scale, step schedule, numerical precision, backend pathway, finite seed set, and near-exact tolerance policy. Nevertheless, the artifact-based workflow provides a reproducible basis for further investigation. The complete software codebase and generated results are made available at

\url{https://github.com/fatahjamro/amub-optimization-software} and archived on Zenodo at \url{https://doi.org/10.5281/zenodo.20689038}.

Future work includes larger seed campaigns, additional optimizer and parameterization comparisons, systematic \(n=7\) campaigns, extended precision studies, validation in further prime and prime-power dimensions, and broader searches in other composite dimensions.

\section*{Acknowledgments}

This work was supported by Atlantic Technological University, Ireland, under the Postgraduate Research Training Programme in Modelling and Computation for Health and Society (MOCHAS-PRTP). We acknowledge the ATU JANUS High-Performance Computing (HPC) cluster facility for providing computational resources used in the multi-seed campaigns and hardware benchmarks. We thank \href{https://pure.atu.ie/en/persons/michael-mccann-3/}{Dr. Michael McCann}, Director of the JANUS Research Centre, for facilitating access to the JANUS HPC resources.

\bibliographystyle{quantum}
\bibliography{bibliography}

\begin{thebibliography}{10}

\bibitem{klappenecker2004ConstructionsMutuallyUnbiased}
Andreas Klappenecker and Martin R{\"o}tteler.
\newblock ``Constructions of mutually unbiased bases''.
\newblock In Finite Fields and Applications: 7th International Conference, Fq7, Toulouse, France, May 5-9, 2003. Revised Papers.
\newblock \href{https://dx.doi.org/10.1007/978-3-540-24633-6_10}{Pages 137--144}.
\newblock Springer~(2004).

\bibitem{bennett2014QuantumCryptographyPublic}
Charles~H. Bennett and Gilles Brassard.
\newblock ``Quantum cryptography: Public key distribution and coin tossing''.
\newblock \href{https://dx.doi.org/10.1016/j.tcs.2014.05.025}{Theoretical Computer Science {\bf 560}, 7--11}~(2014).

\bibitem{WOOTTERS1989optimalStateMUB}
William~K Wootters and Brian~D Fields.
\newblock ``Optimal state-determination by mutually unbiased measurements''.
\newblock \href{https://dx.doi.org/10.1016/0003-4916(89)90322-9}{Annals of Physics {\bf 191}, 363--381}~(1989).

\bibitem{McNulty2026mutuallyunbiased}
Daniel McNulty and Stefan Weigert.
\newblock ``Mutually {U}nbiased {B}ases in {C}omposite {D}imensions - {A} {R}eview''.
\newblock \href{https://dx.doi.org/10.22331/q-2026-04-01-2051}{{Quantum} {\bf 10}, 2051}~(2026).

\bibitem{bengtsson2007MUBDandHadamard}
Ingemar Bengtsson, Wojciech Bruzda, Åsa Ericsson, Jan-Åke Larsson, Wojciech Tadej, and Karol Życzkowski.
\newblock ``Mutually unbiased bases and hadamard matrices of order six''.
\newblock \href{https://dx.doi.org/10.1063/1.2716990}{Journal of Mathematical Physics {\bf 48}, 052106}~(2007).

\bibitem{horodecki2022FiveOpenProblems}
Pawe\l{} Horodecki, \L{}ukasz Rudnicki, and Karol \ifmmode~\dot{Z}\else \.{Z}\fi{}yczkowski.
\newblock ``Five open problems in quantum information theory''.
\newblock \href{https://dx.doi.org/10.1103/PRXQuantum.3.010101}{PRX Quantum {\bf 3}, 010101}~(2022).

\bibitem{brierley2010mutuallyunbiasedbasesdimensions}
Stephen Brierley, Stefan Weigert, and Ingemar Bengtsson.
\newblock ``All mutually unbiased bases in dimensions two to five''.
\newblock \href{https://dx.doi.org/10.26421/QIC10.9-10-6}{Quantum Information \& Computation {\bf 10}, 803--820}~(2010).

\bibitem{elena2000ApproxLieAlgerba}
Elena Celledoni and Arieh Iserles.
\newblock ``Approximating the exponential from a lie algebra to a lie group''.
\newblock \href{https://dx.doi.org/10.1090/S0025-5718-00-01223-0}{Math. Comput. {\bf 69}, 1457--1480}~(2000).

\bibitem{helmberg2008introHilbertspace}
Gilbert Helmberg.
\newblock ``Introduction to spectral theory in hilbert space''.
\newblock Courier Dover Publications. ~(2008).
\newblock  url:~\url{https://www.sciencedirect.com/science/book/9780720423563}.

\bibitem{littlewood1977theoryofgroupandmatrix}
Dudley~Ernest Littlewood.
\newblock ``The theory of group characters and matrix representations of groups''.
\newblock \href{https://dx.doi.org/10.1090/chel/357}{Volume 357}.
\newblock American Mathematical Soc. ~(1977).

\bibitem{durt2010OnMutuallyUnbiasedBases}
THOMAS DURT, BERTHOLD-GEORG ENGLERT, INGEMAR BENGTSSON, and KAROL \.{Z}YCZKOWSKI.
\newblock ``On mutually unbiased bases''.
\newblock \href{https://dx.doi.org/10.1142/S0219749910006502}{International Journal of Quantum Information {\bf 08}, 535--640}~(2010).

\bibitem{brierley2009MutuallyUnbiasedBases}
Stephen Brierley.
\newblock ``Mutually unbiased bases in low dimensions''.
\newblock PhD thesis.
\newblock University of York.
\newblock ~(2009).
\newblock  url:~\url{https://etheses.whiterose.ac.uk/id/eprint/587/}.

\bibitem{Grassl2009OnSICPOVMsMUBsd6}
Markus Grassl.
\newblock ``{On SIC-POVMs and MUBs in Dimension 6}''~(2004).
\newblock  \href{http://arxiv.org/abs/quant-ph/0406175}{arXiv:quant-ph/0406175}.

\bibitem{Paszke2017AutomaticDI}
Adam Paszke, Sam Gross, Soumith Chintala, Gregory Chanan, Edward Yang, Zachary DeVito, Zeming Lin, Alban Desmaison, Luca Antiga, and Adam Lerer.
\newblock ``Automatic differentiation in pytorch''.
\newblock In NIPS 2017 Workshop on Autodiff.
\newblock ~(2017).
\newblock  url:~\url{https://openreview.net/forum?id=BJJsrmfCZ}.

\bibitem{pytorch2019AnImperativeStyle}
Adam Paszke, Sam Gross, Francisco Massa, Adam Lerer, James Bradbury, Gregory Chanan, Trevor Killeen, Zeming Lin, Natalia Gimelshein, Luca Antiga, Alban Desmaison, Andreas Kopf, Edward Yang, Zachary DeVito, Martin Raison, Alykhan Tejani, Sasank Chilamkurthy, Benoit Steiner, Lu~Fang, Junjie Bai, and Soumith Chintala.
\newblock ``Pytorch: An imperative style, high-performance deep learning library''.
\newblock In H.~Wallach, H.~Larochelle, A.~Beygelzimer, F.~d\textquotesingle Alch\'{e}-Buc, E.~Fox, and R.~Garnett, editors, Advances in Neural Information Processing Systems 32.
\newblock \href{https://dx.doi.org/10.48550/arXiv.1912.01703}{Pages 8024--8035}.
\newblock Curran Associates, Inc.~(2019).

\bibitem{Higham2008FunctionsMatrix}
Nicholas~J. Higham.
\newblock ``Functions of matrices''.
\newblock \href{https://dx.doi.org/10.1137/1.9780898717778}{Society for Industrial and Applied Mathematics}. ~(2008).

\bibitem{musto2019QuantumLatinSquares}
B.~J. Musto.
\newblock ``Quantum latin squares and quantum functions: Applications in quantum information''.
\newblock PhD thesis.
\newblock University of Oxford.
\newblock ~(2019).
\newblock  url:~\url{https://www.cs.ox.ac.uk/people/bob.coecke/BenMustoThesis}.

\bibitem{knill1996NonBinaryUnitary}
Emanuel Knill.
\newblock ``Non-binary unitary error bases and quantum codes''~(1996).
\newblock  \href{http://arxiv.org/abs/quant-ph/9608048}{arXiv:quant-ph/9608048}.

\bibitem{Vicary_2019BiunitaryConstructions}
David~J. Reutter and Jamie Vicary.
\newblock ``Biunitary constructions in quantum information''.
\newblock \href{https://dx.doi.org/10.21136/hs.2019.04}{Higher Structures {\bf 3}, 109--154}~(2019).

\bibitem{IBMQuantumMarrakeshBackend}
{IBM Quantum}.
\newblock ``{IBM Quantum Platform Compute Resources: ibm\_marrakesh}''.
\newblock \url{https://quantum.cloud.ibm.com/computers?system=ibm_marrakesh}~(2026).
\newblock Accessed 2026-06-14.

\end{thebibliography}

\clearpage
\onecolumn
\section{Appendix: Additional Figures}

\begin{figure}[htbp]
  \centering
  \includegraphics[width=0.95\textwidth]{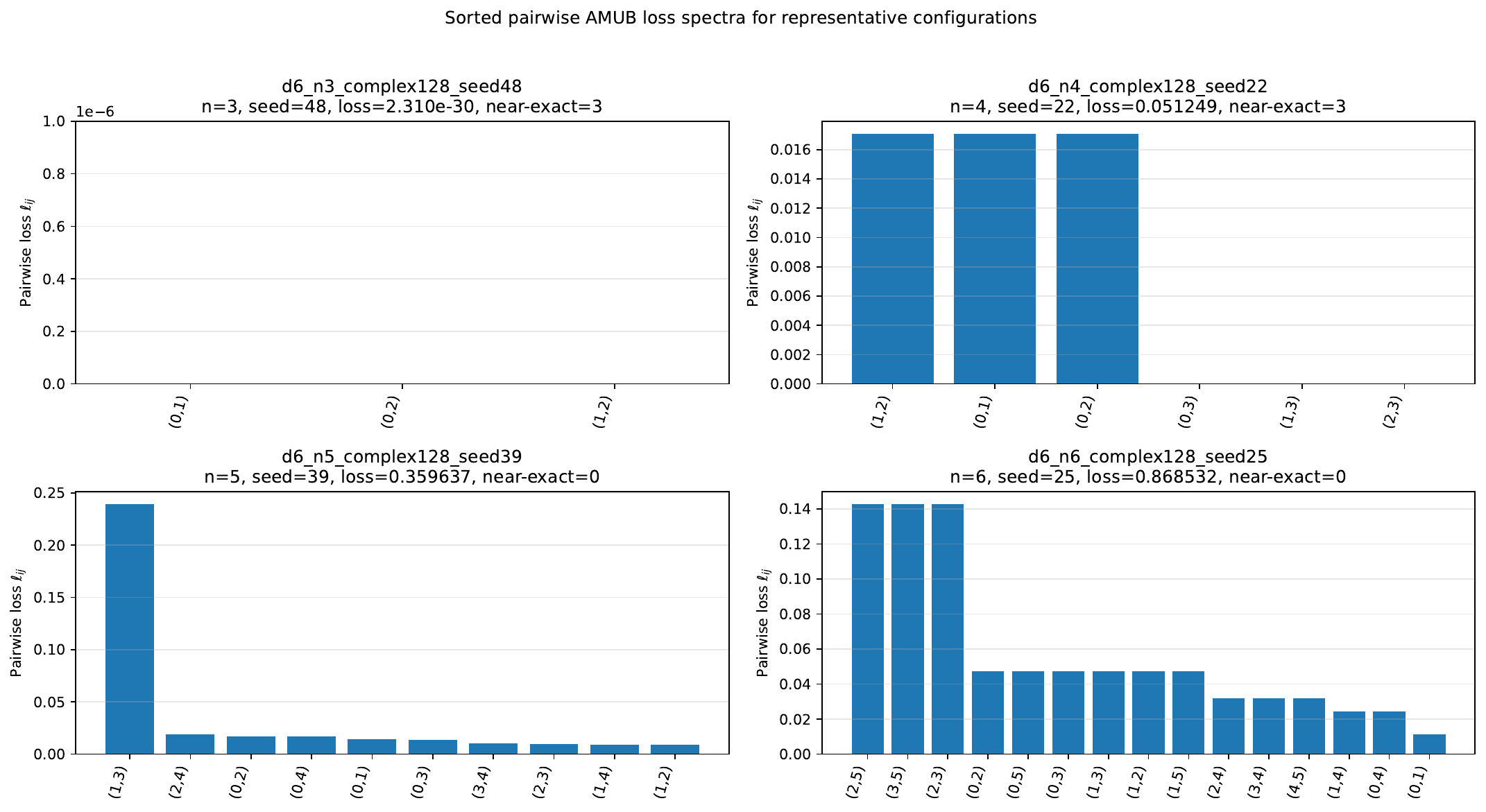}
  \caption{Sorted pairwise AMUB loss spectra for representative complex128 configurations. Each bar corresponds to one pairwise contribution $\ell_{ij}$. The exact $n=3$ representative has pairwise losses at numerical zero, the defective $n=3$ representative has three approximately equal nonzero losses, the $n=4$ representative exhibits three near-zero losses and three equal defective losses, and the $n=5$ and $n=6$ representatives have no zero pairwise losses.}
  \Description{A multi-panel bar chart showing sorted pairwise AMUB loss values for representative complex128 configurations. The exact n equals 3 panel has zero bars, the defective n equals 3 panel has three equal nonzero bars, the n equals 4 panel has three near-zero bars and three equal nonzero bars, and the n equals 5 and n equals 6 panels have all nonzero bars.}
  \label{fig:pairwise-loss-spectra-complex128}
\end{figure}

\end{document}